\documentclass[12pt]{iopart}
\usepackage[T1]{fontenc}
\usepackage{graphicx}

\usepackage{xcolor}

\usepackage{ulem}

\def \sriro{Sr${}_2$IrO$_{4}$}

\def \t2g{t$_{2g}$}

\def \dx2y2{d$_{x^2-y^2}$}
\def \j32{$j_{\mathrm{eff}}$=$3/2$}
\def \jeff12{$j_{\mathrm{eff}}$=$1/2$}

\begin{document}
\article{JPCM Topical Review}{Spectral functions of Sr${}_2$IrO${}_4$: Theory versus experiment}
\author{B Lenz$^1$, C Martins$^2$ and S Biermann$^{1,3}$}
\address{$^1$ CPHT, Ecole Polytechnique, CNRS, Universit{\'e} Paris-Saclay, Route de Saclay, 91128 Palaiseau, France}
\address{$^2$ Laboratoire de Chimie et Physique Quantiques, UMR 5626, Universit{\'e} Paul Sabatier, 118 route de Narbonne, 31400 Toulouse, France}
\address{$^3$ Coll{\`e}ge de France, 11 Place Marcelin Berthelot, 75005 Paris, France}
\ead{silke.biermann@polytechnique.edu}

%%%%%%%%%%%%%%%%%

\begin{abstract}
The spin-orbit Mott insulator \sriro\ has attracted a lot of interest in recent years from theory and experiment due to its close connection to isostructural high-temperature copper oxide superconductors.
Despite of not being superconducting its spectral features closely resemble those of the cuprates, including Fermi surface and pseudogap properties.
In this article, we review and extend recent work in the theoretical description of the spectral function of pure and electron-doped \sriro\ based on a cluster extension of dynamical mean-field theory (''oriented-cluster DMFT``) and compare it to available angle-resolved photoemission data.
Current theories provide surprisingly good agreement for pure and electron-doped \sriro, both in the paramagnetic and antiferromagnetic phases. 
Most notably, one obtains simple explanations for the experimentally observed steep feature around the $M$ point and the pseudo-gap-like spectral feature in electron-doped \sriro.

\end{abstract}
\submitto{\JPCM}
\maketitle

%%%%%%%%%%%%%%%%%%%%%%%%%%%%%%%%%%%%%%%%%%%%%%%%%
\section{Introduction}

Electronic Coulomb correlations are well-established as one of the main players determining the excitations and response properties of 3d transition metal oxides or f-electron compounds. 
The relatively modest spatial extension of the orbitals forming the valence states in these materials leads to comparably large effective local Coulomb interactions. 
They are in particular large enough to compete with the kinetic energy even when several partially filled valence states are involved.
It is well-known from studies of the multi-orbital Hubbard model that the critical interaction strength needed to localize the electrons and induce a Mott insulating state grows with the orbital degeneracy \cite{Gunnarsson1996,Rozenberg1997}, translating the enhanced kinetic energy in the multi-orbital context.

Multi-orbital physics is rather the generic situation in transition metal oxides \cite{Imada1998} and even more so in f-electron compounds. 
Examples include t$_{2g}$-systems such as titanates \cite{Pavarini2004} or vanadates \cite{Tomczak2014}, or Mn-, Fe- or Ni-based e$_g$-oxides \cite{Seth2017}.
A notable exception are high-temperature superconducting cuprates, where a single hole in an otherwise completely filled 3d-shell can be dominantly attributed to an orbital of d$_{x^2-y^2}$ character, corresponding -- in the parent compounds -- to a nominally half-filled configuration.
Most of the exotic properties of cuprates are directly or indirectly attributed to this specific situation \cite{Lee2006,Scalapino2012,Alloul2014} \footnote{However, it should be noted that such a one-band minimal model has been repeatedly questioned to reproduce all eminent features of cuprate physics \cite{Andersen1995,Hanke2010}.}.

Iridium oxides have added interesting twists to these quite general considerations \cite{Kim2008}.
First, the rather extended nature of the Ir 5d orbitals would suggest correlations to play a minor role in these systems if indeed all three t$_{2g}$ orbitals were contributing to the low-energy physics around the Fermi level in a more or less degenerate way.
The effective local Coulomb interactions have rather modest values of the order of $2$ eV \cite{Martins2011, Arita2012, Martins2017}, comparable to the t$_{2g}$-electron bandwidth, and would -- in a degenerate three-orbital model -- at most lead to a modest mass enhancement in a metallic state \cite{Martins2011}. 
A radically different situation is realized in nature: focussing on the prototypical Sr$_2$IrO$_4$, one is dealing with an insulating material with a strongly temperature-dependent gap, both in its low-temperature magnetic and high-temperature paramagnetic phase, as confirmed by x-ray photoelectron \cite{Haskel2012,Fujiyama2012,Kim2012,Fujiyama2014, Kim2014}, angle-resolved photoemission \cite{Kim2008,Wang2013,Kim2014,Yamasaki2014,Brouet2015,Torre2015,Liu2015,Nie2015,Piovera2016,Cao2016,Kim2016,Martins2018}, and scanning tunneling \cite{Li2013, Dai2014} spectroscopy, optical conductivity measurements \cite{Moon2008,Kim2008,Moon2009,Hsieh2012,Zhao2016,Proepper2016,Li2016}
as well as magneto \cite{Chikara2009, Ge2011}, thermo \cite{Pallecchi2016}, and electrical \cite{Klein2008,Kim2009,Zhou2017,Zocco2014, Fruchter2015} transport measurements.

From the theoretical point of view, \sriro\ has been investigated by density functional theory in combination with single-site dynamical mean-field theory \cite{Martins2011,Arita2012,Zhang2013,Li2013,Yamasaki2014} or its cluster extensions \cite{Hampel2015, Martins2018,Moutenet2018}, the self-consistent Born approximation \cite{Paerschke2017} or the self-consistent Hartree-Fock method \cite{Zhou2017b}.

The system does not explore the full space of orbital fluctuations corresponding to the nominal t$_{2g}^5$ configuration, but rather adopts cuprate-like single orbital physics.
The microscopic reason can be found in a subtle interplay between spin-orbit interactions, structural distortions and Coulomb correlations \cite{Martins2011,Martins2017}.
The former two result in a rearrangement of the electronic states in such a way that -- at the one-particle level -- two nearly filled bands (usually labeled by an effective quantum number $j_{eff}=3/2$) and a half-filled ($j_{eff}=1/2$) band can be identified.
The role of the Coulomb correlations is then two-fold: first, to enhance the spin-orbital polarization of this state such as to leave only the now exactly half-filled $j_{eff}=1/2$ at the Fermi level, and second, to localize the hole in this state.

\begin{figure}[h]
	\includegraphics[width=\linewidth]{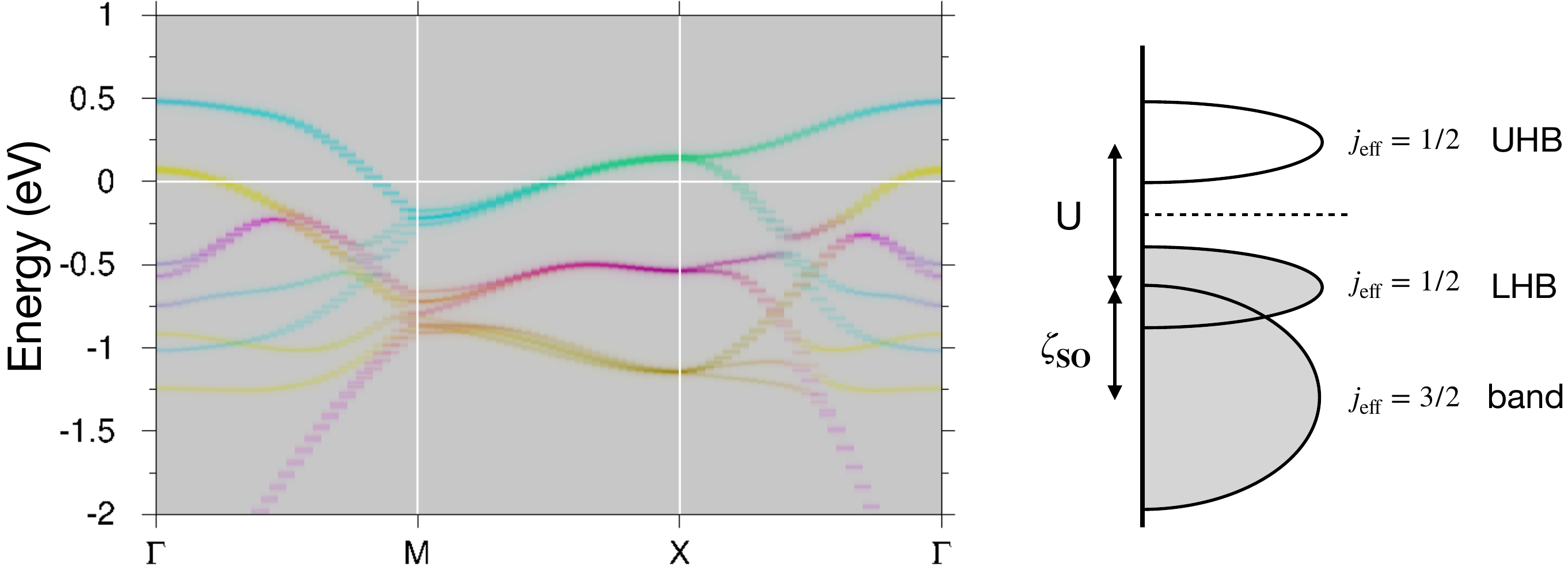}
	\caption{\label{Fig:Intro2}Left: Wannier-projected \jeff12 (turquoise) and \j32 bands ($\vert m_j\vert=1/2$ in magenta, $\vert m_j\vert=3/2$ in yellow) within DFT+SOC. 
		Right:  After applying Coulomb interactions, the \jeff12 manifold splits into an upper 
		and lower Hubbard band. Level scheme adopted from \textit{Kim et al. 2008} \cite{Kim2008}.}
\end{figure}

The net result is thus an insulating state that bears some resemblance to the one of the parent compounds of the cuprates: it is realized within a layered crystal structure that deviates from the one of the celebrated La$_2$CuO$_4$ cuprate only by rotations of the oxygen octahedra around the crystallographic c-axis; it is of single-orbital nature, and it is accompanied by antiferromagnetic order in the low-temperature regime.
It has been noted that Sr$_2$IrO$_4$ could be described by a  twisted Hubbard model \cite{Senthil2011}, bridging to the most popular model for high-temperature superconducting cuprates.
In both cases, the antiferromagnetic order is driven by superexchange constants, which are of the same order of magnitude for \sriro\ (nearest neighbor exchange $J_{NN}$ between $\sim 0.06$eV \cite{Gretarsson2016,Pincini2017} and $\sim0.1$eV \cite{Fujiyama2012}) and its related cuprate La${}_2$CuO${}_4$ ($J_{NN}\sim0.14$eV \cite{Headings2010}). 
In Sr$_2$IrO$_4$, a canted antiferromagnetic ground state is formed below the N\'eel temperature of $T_N\sim240$K, which amounts to a weak ferromagnetic moment in $c$-direction \cite{Ye2013}.

Doping by chemical substitutions or vacancies leads in the cuprates to by now well-characterized but still intriguing superconducting phases. 
In iridates, doping has been achieved by 
La- \cite{Cosio-Castaneda2007,Klein2008},
Rh- \cite{Klein2008}, 
or Co-substitutions \cite{Gatimu2012}, 
O-vacancies \cite{Korneta2010}
and surface-doping \cite{Kim2014,Kim2016}. 
From those, in particular Lanthanum and Rhodium substitutions have been used in recent years to systematically study electron- and hole-doped \sriro\ respectively.
In both cases, the substitutions primarily act like electronic doping and do not change much the lattice constant of \sriro\ \cite{Klein2008,Cosio-Castaneda2007}.
Yet, any direct observation of superconductivity remains elusive.
In the cuprates, the normal state is itself anything but normal: the most intriguing feature is probably the appearance of a pseudogap in the low-energy electronic states in immediate vicinity to the Fermi level \cite{Alloul1989}, which has continued to stimulate experimental \cite{Taillefer2010,Sacuto2013,Kordyuk2015, Proust2018} and theoretical \cite{Phillips2003,Lee2006,Scalapino2012} studies.

Over the last decades, spectroscopic techniques have evolved into a most powerful tool to characterize low-energy excitations in correlated materials in a direct fashion (for a review, see e.g. \cite{vanRoekeghem2016}). 
Up to matrix elements translating modulations induced by light-matter coupling, angle-resolved photoemission measures the spectral function $A(k,\omega)$ related to the one-particle Green's function by
\begin{eqnarray}
	A(k,\omega) = - \frac{1}{\pi}\ \Tr\ \mathrm{Im}\ G(k, \omega).
\end{eqnarray}
The latter is in principle accessible by advanced many-body techniques, such as dynamical mean field theory (for reviews see \cite{Georges1996, Kotliar2004}) and its extensions. 
Recent advances using combined density functional theory dynamical mean field schemes \cite{Anisimov1997,Lichtenstein1998, Biermann2014} have made substantial progress in describing spectral properties of correlated materials in an {\it ab initio} (i.e. materials-specific) fashion (see e.g. \cite{Biermann2005, Kotliar2006,Ma2014, Nowadnick2015, Hausoel2017, Hariki2017, Mandal2018}). 
Iridates have been in the focus since 2010 \cite{Martins2010, Martins2011, Arita2012, Hampel2015, Zhang2013}, with most recent advances giving a detailed account of spectral properties both in the pure and doped samples based on an extension of DMFT \cite{Martins2018}.

In this paper, we review and extend recent theoretical results using the cluster extension of DMFT proposed in Ref.~\cite{Martins2018} in comparison to available experimental data to assess the current state of the art in the field.

\begin{figure}[h]
	\includegraphics[width=\linewidth]{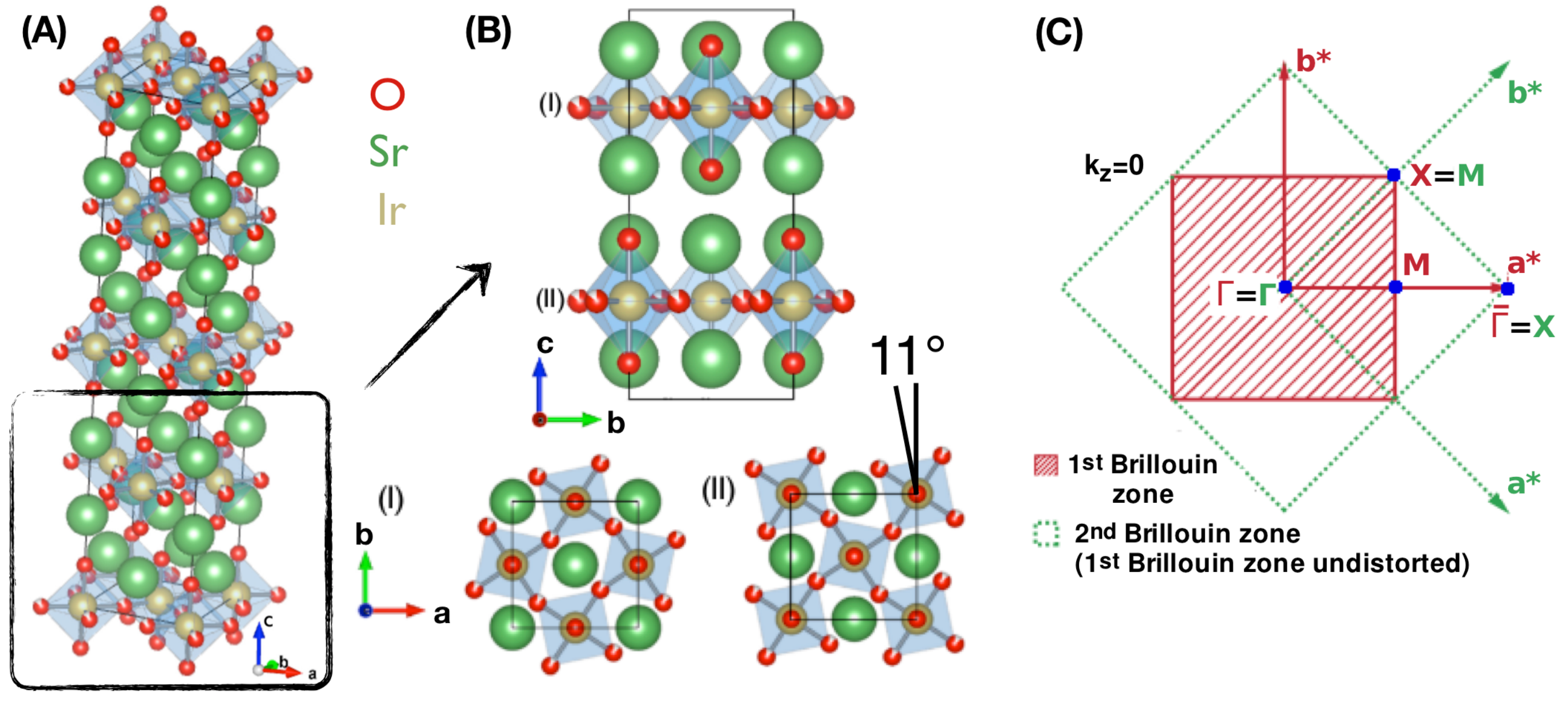}
	\caption{\label{Fig:Intro1}Crystal structure of \sriro. 
	\sriro\ crystallizes in a $I4_1/acd$ structure (A) and can be described by a unit cell which comprises two Sr${}_2$IrO${}_4$ layers (I) and (II) in c-direction (B) with a total of four Ir atoms (two per layer). 
	Within such a layer, the environments of neighboring Ir atoms differ since the surrounding oxygen octahedra are rotated by $\sim11^{\circ}$ in opposite directions. 
	Due to these rotations of the oxygen octahedra the unit cell is doubled and the respective Brillouin zone takes only half of the size of an undistorted geometry (C). Panels (B) and (C) adopted from \textit{Martins et al.} \cite{Martins2018}.}
\end{figure}

The paper is organized as follows:
In the next section the oriented cluster dynamical mean-field theory of Ref.~\cite{Martins2018} is reviewed.
Sections~\ref{Sec:PM} and~\ref{Sec:AF} focus on the paramagnetic high-temperature and antiferromagnetic low-temperature phase of undoped \sriro\ respectively.
We compare calculated spectra to experimentally measured ones in both cases.
Electron-doped \sriro\ is discussed in Sec.~\ref{Sec:ElDop} with a focus on Fermi surface and pseudogap features of the spectral function.
A summary and outlook conclude the paper in Sec.\ref{Sec:Out}.

%%%%%%%%%%%%%%%%%%%%%%%%%%%%%%%%%%%%%%%%%%%%%%%%%
\section{Technique: Oriented Cluster Dynamical Mean-Field Theory}
In dynamical mean-field theory (DMFT) \cite{Georges1996, Kotliar2004} the interacting many-electron problem is mapped onto a single impurity site that is coupled dynamically to a non-interacting bath.
Since this impurity represents an arbitrary site in the lattice it amounts to a purely local description and correlation effects, which are taken into account via the impurity self-energy $\Sigma_{\mathrm{imp}}$, are therefore local in nature, too.
The momentum-dependent lattice Green's function is constructed using $\Sigma_{\mathrm{imp}}$ via
\begin{eqnarray}
\label{Eq:Greens}
G(\mathbf{k},\omega)=\left[\omega+\mu-H(\mathbf{k})-\Sigma_{\mathrm{imp}}(\omega)\right]^{-1}.
\end{eqnarray} 
In the case of \sriro, the one-particle hamiltonian $H(\mathbf{k})$ is obtained from a tight-binding fit of the \jeff12 band, see \ref{App:TB}.
The Green's function of the impurity problem is then equal to the local Green's function of the lattice problem, which is ensured by the DMFT self-consistency condition
\begin{eqnarray}
	\label{Eq:SCC}
	G_{\mathrm{loc}}(\omega)=\sum_{\mathbf{k}}G(\mathbf{k},\omega),
\end{eqnarray} 
which determines the bath and restores the translational invariance.
In general, the system is of multi-orbital nature with multiple correlated shells $a$ per unit cell. 
For each correlated site, the self-energy of the effective impurity site has to be rotated into the local reference frame via
\begin{eqnarray}
	\Sigma_a(\omega) = R^{\dagger}_a\Sigma_{\mathrm{imp}}(\omega)R_a.
\end{eqnarray}
Here, $\Sigma_{\mathrm{imp}}$ is a matrix in orbital space, i.e. a $N\times N$ matrix for a system consisting of $N$ correlated orbitals per correlated shell. 
If we assume that --due to crystallographic symmetries-- the $N_a$ correlated sites are equivalent, the overall self-energy matrix $\Sigma$ is of size $N\cdot N_a\times N\cdot N_a$. 
Moreover, the self-energy is block-diagonal in the correlated shells, i.e.
\begin{equation}
	\Sigma_{aa^{\prime}}(\omega) = \delta_{aa^{\prime}}\Sigma_a(\omega).
\end{equation}
Since all shells are rotated into their local coordinate system, using the self-energy $\Sigma_{aa^{\prime}}(\omega)$ in Eq.(\ref{Eq:Greens}) leads to a lattice Green's function, which respects all point group symmetries of the crystal.
\begin{figure}[h]
\begin{center}
\includegraphics[width=.9\linewidth]{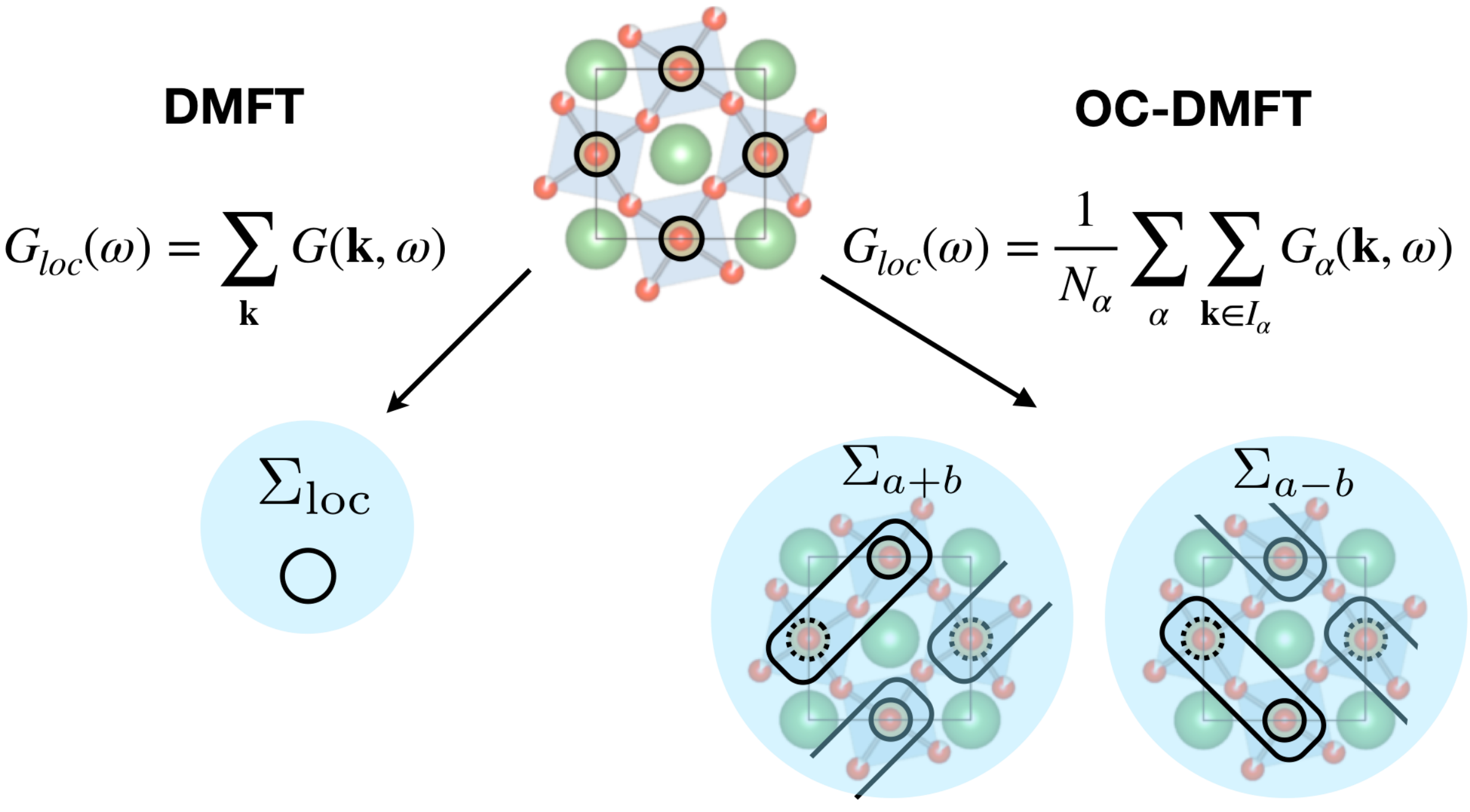}
\caption{\label{Fig:Tech}Comparison of DMFT and OC-DMFT: Since more than one cluster orientation is allowed, the self-consistency condition includes an orientation average which restores the point-group symmetry. In this case, the momenta $\mathbf{k}$ have to be taken out of the Brillouin zone $I_{\alpha}$ corresponding to the respective orientation.}
\end{center}
\end{figure}

In cluster extensions of DMFT \cite{Maier2005} the original lattice is tiled into (identical) clusters, which can break the point group symmetry of the crystal.
The location $\mathbf{x}$ of each site can then be represented by specifying the origin of a cluster $\mathbf{\tilde{x}}$ and the site within the cluster $\mathbf{X}$.
Using the self-consistency condition of Eq.(\ref{Eq:SCC}) only restores the translational invariance, but does not remove possible violations of the point group symmetry.
Instead of limiting ourselves to clusters, which have the full point group symmetry of the original lattice, we recently proposed to use oriented clusters and to restore the point group symmetry via a modified embedding condition \cite{Martins2018}.
Choosing an oriented two-atom dimer as a cluster, which represents two nearest-neighbor Ir sites, one is left with two possible (perpendicular) cluster orientations. 
Therefore, we can define a momentum-resolved Green's function evaluated with an orientation dependent self-energy:
\begin{equation}
	G_{\alpha}(\mathbf{K},\omega) = \left[\omega + \mu - H(\mathbf{K}) - \Sigma^{\alpha}_{\mathrm{dim}}\right]^{-1}.
\end{equation}
The momenta $\mathbf{K}$ are taken out of the Brillouin zone corresponding to the Bravais lattice of the cluster locations $\mathbf{\tilde{x}}$ \cite{Maier2005}.
As a consequence, when constructing the local Green's function, the sum now runs over the two possible orientations $\alpha=x,y$ and over their corresponding momenta $\mathbf{K}\in I_{\alpha}$:
\begin{equation}
	\label{Eq:SCE}
	G_{\mathrm{loc}}(\omega)=\frac{1}{N_{\alpha}}\sum_{\alpha}\sum_{\mathbf{K}\in I_{\alpha}}G_{\alpha}(\mathbf{K},\omega).
\end{equation}
The momentum-resolved spectral function $A(\mathbf{k},\omega)$ to be compared with the ARPES spectra later on is obtained from the k-resolved Green function
\begin{equation}
	\label{Eq:Gk}
	G(\mathbf{k},\omega)=\frac{1}{N_{\alpha}}\sum_{\alpha}G_{\alpha}(\mathbf{k},\omega).
\end{equation}
Here, $\mathbf{k}$ is the momentum-vector in the crystallographic Brillouin zone of the lattice. 

For \sriro\ neighboring Ir sites already differ in the rotation of their oxygen octahedra and the crystallographic unit cell consists of four Ir atoms.
A possible choice for the unit cell amounts to taking two neighboring Ir sites in two stacked a-b layers in c-direction, see Fig.~\ref{Fig:Intro1}(A).
Choosing two dimers as the building block for a cluster tiling of the lattice therefore does not change the Brillouin zone and keeps the same crystallographic cell.

Since we use the minimal cluster that incorporates non-local effects, it is possible to gain insight into the physical processes which cause certain features in the spectrum.
For instance, inter-site fluctuations of the dimer are of antiferromagnetic spin nature and are encoded in the non-local part of the cluster self-energy.
The latter can be seen as energy-dependent modulations of the inter-site hopping processes that enter in the Green's function $G_{\alpha}$.
In \sriro\ this selection of an orientation $\alpha$ causes a $x-y$-symmetry breaking in $G_{\alpha}$. This shows the importance of the orientational average in Eq.~(\ref{Eq:SCE}), which restores this symmetry.
Most interestingly, signatures of the fluctuations can be seen in terms of tilted lens-shaped Fermi surfaces in electron-doped \sriro\ as discussed later in Sec.~\ref{Sec:ElDop}.

\begin{figure}[h]
\includegraphics[width=\linewidth]{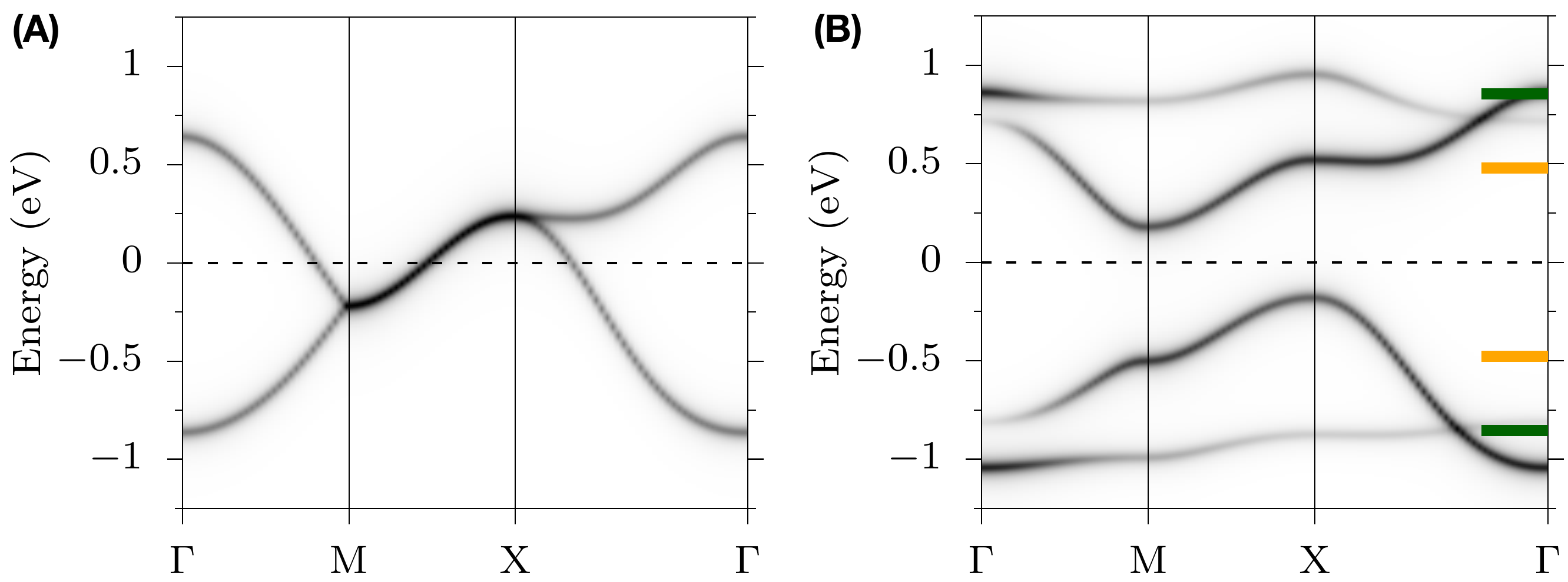}
\caption{\label{Fig:Tech2}(A) The dispersion of one of the two downfolded inter-layer tight-binding bands constructed in \ref{App:TB} with chemical potential adjusted to match half-filling. 
(B) Effect of treating the Hubbard interaction $U=1.1eV$ in Hubbard-I approximation via OC-DMFT. The four poles of the dimer Green's function are indicated by orange (green) lines and correspond to the bonding-/antibonding (satellite) bands.}
\end{figure}

Figure~\ref{Fig:Tech2} illustrates schematically how including non-local fluctuations via the dimer self-energy $\Sigma_{\alpha}$ within OC-DMFT affects the spectral function.
The starting point of the calculations is a tight-binding fit of the DFT \jeff12 bands, which defines the bare dispersion $\epsilon(\mathbf{k})$.
Since the unit cell comprises two layers in c-direction with a dimer each, we take the weak inter-layer hopping into account and obtain two slightly different downfolded inter-layer bands as discussed in \ref{App:TB}.
For each of the two dispersions we then calculate the lattice Green's function for the two possible dimer orientations. 
Choosing one dimer configuration changes the non-interacting spectrum of the band shown in Fig.~\ref{Fig:Tech2}(A) into the interacting spectral function $A_{\alpha}(\mathbf{k},\omega)$ shown in (B).
Besides the splitting of the band into a bonding and an antibonding band, additional satellite bands emerge at larger energy.
For comparison, the bonding and antibonding poles of the dimer Green's function are indicated by orange, the position of their satellites by green lines.

We construct a Hamiltonian to model the \jeff12 band which amounts to a one-band Hubbard model of interaction strength $U_{\mathrm{eff}}=1.1$eV using a tight-binding modelling of the dispersion.
This effective interaction $U_{\mathrm{eff}}$ takes into account the large nearest-neighbor interaction, which is due to the extended nature of the $5d$ orbitals.
On the level of our dimer solver, the value of $1.1$eV is in good agreement with recent constrained random phase approximation (cRPA) \cite{Aryasetiawan2004} estimates \cite{Martins2011, Backes2019}\footnote{For a dimer, the nearest-neighbor interaction $V$ reduces the on-site interaction $U$ to $U_{\mathrm{eff}}=U-V$. After averaging the cRPA interaction values over the three $t_{2g}$ orbitals, one arrives at values of $U_{\mathrm{eff}}=1.3$eV \cite{Martins2011} or $U_{\mathrm{eff}}=1.0$eV \cite{Backes2019}.} 
and produces a gap size of $\Delta\approx0.3$eV consistent with experiment \cite{Moon2009}.
Furthermore, if we assume strong coupling and do a rough estimation of the exchange coupling via $J=4t^{2}/U_{eff}$, using the two different dispersions we obtain values of $J=0.09$eV and $J=0.13$eV,  which are close to the values determined from experiment \cite{Gretarsson2016,Pincini2017,Fujiyama2012}.

To solve the cluster problem, we generalized to a two-site cluster the Hubbard-I solver \cite{Hubbard1963}.
Hubbard-I consists in neglecting the hybridization in the calculation of the self-energy and is therefore appropriate for materials in the strong-coupling limit.
For this reason it has for example been used for rare-earth elements and materials \cite{Lichtenstein1998,Locht2017}.
Here, our philosophy is different: We will show in the following that the dimer Hubbard-I approximation gives an excellent account of the spectral properties of \sriro\ despite its proximity to the Mott transition.
For more details on the technique itself, the cluster Hubbard-I solver used for the calculations and how to determine a thermodynamically consistent charge, we refer the reader to the supplementary material of Ref.~\cite{Martins2018} and references therein.
Further details on the Hubbard dimer can be found in Ref.~\cite{Tomczak2007}.

%%%%%%%%%%%%%%%%%%%%%%%%%%%%%%%%%%%%%%%%%%%%%%%%
\section{\sriro\ in the Paramagnetic High-Temperature Phase}
\label{Sec:PM}
At high temperatures, \sriro\ is a paramagnetic spin-orbit Mott insulator \cite{Kim2008}. 
In the following section we discuss the spectral function in this paramagnetic phase, both calculated from OC-DMFT and measured using ARPES.

%%%%%%%%%%%%%%%
\subsection{Spectra from Theory}

\begin{figure}[h!]
	\includegraphics[width=.9\linewidth]{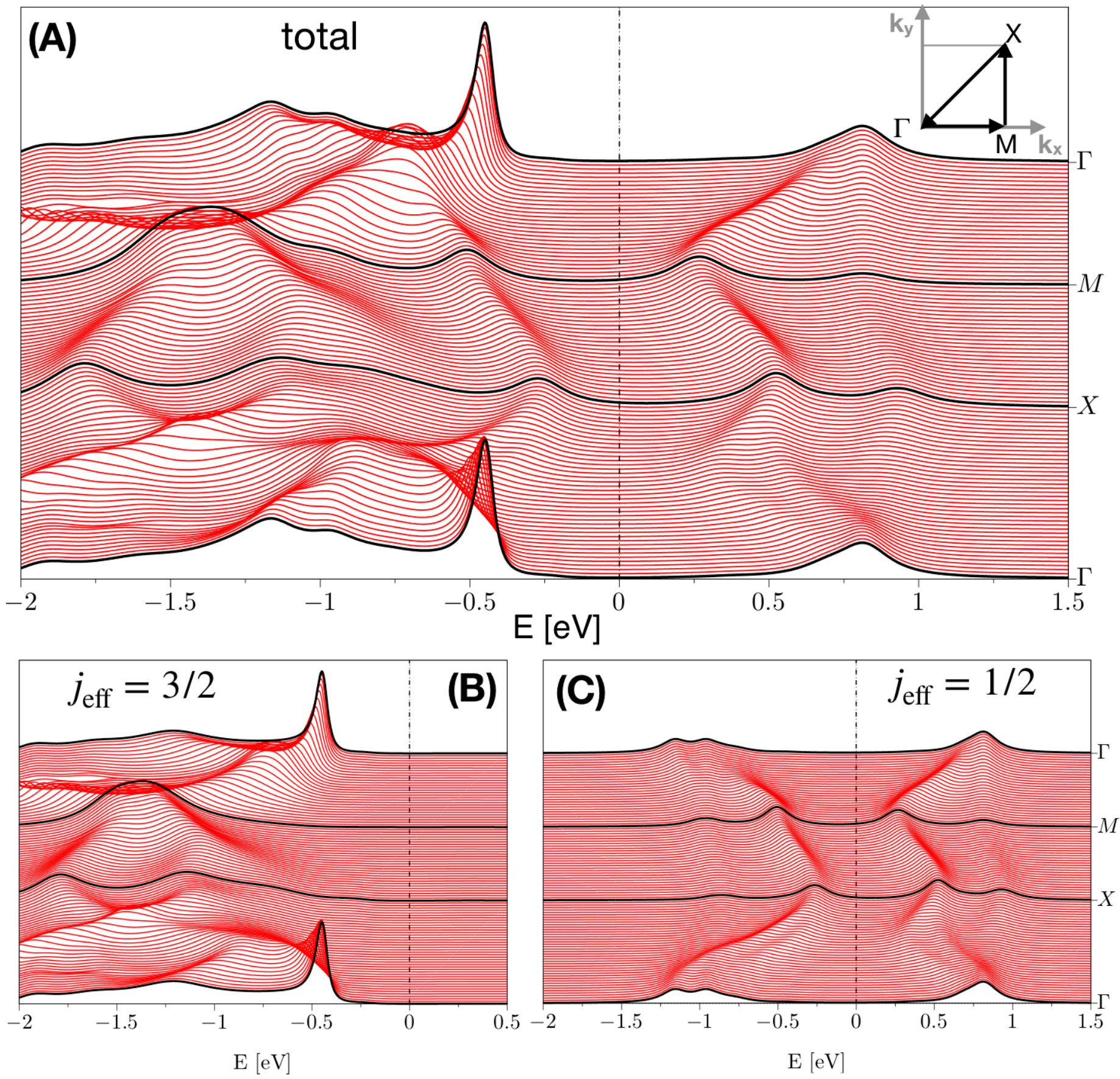}
\caption{\label{Fig:PMSpec_1}{Calculated spectral function $A(\mathbf{k},\omega)$ of pure \sriro\ in the paramagnetic high-temperature phase along the $\mathbf{k}$-path $\Gamma-M-X-\Gamma$, which is shown in the inset of (A). The contribution of the filled \j32 bands is shown in (B) and the spectral weight of the half-filled \jeff12 band in (C).}}
\end{figure}

Figure~\ref{Fig:PMSpec_1} shows the spectral function as well as their \j32 and \jeff12 parts along the path $\Gamma-M-X-\Gamma$.
The model shows a gapped spectrum with the highest occupied states residing at the $X$ point, while the lowest unoccupied states are at the $M$ point.

In the \jeff12 manifold, one can identify four separate peaks out of which the bonding and antibonding ones have largest spectral weight.
Furthermore, one sees small satellites in the electron- and hole-spectrum, as can be seen in Fig.~\ref{Fig:PMSpec_1}(C).
The bonding-antibonding splitting of two neighboring iridate layers in $c$-direction is much smaller such that it does not give rise to further splitting into separable peaks.

The \j32 band does not cross the Fermi level and is completely filled.
Still, it is responsible for spectral features at low energies such as the highest occupied states at and close to the $\Gamma$ point. 
Away from the $\Gamma$ point, the \j32 manifold mainly contributes spectral weight at energies $E<-1$eV.

%%%%%%%%%%%%%%%
\subsection{Experimental Spectra}
In angle-resolved photoemission spectroscopy it is difficult to investigate insulators due to the build-up of surface charge.
However, for insulators with a small gap like \sriro\ it is feasible to measure the spectral function at moderate temperatures (in case of \sriro\ without further ado for synchrotron radiation down to $T\sim50$K \cite{Torre2015,Brouet2015,Cao2016}).
Most of the photoemission studies on \sriro\ focussed on the low-temperature phase, which shows antiferromagnetic long-range order of the $J=1/2$ pseudospins.
Up to our knowledge there is only one recent study presenting photoemission spectra in the paramagnetic high-temperature phase \cite{Martins2018}.

\begin{figure}[h]
	\includegraphics[width = \linewidth]{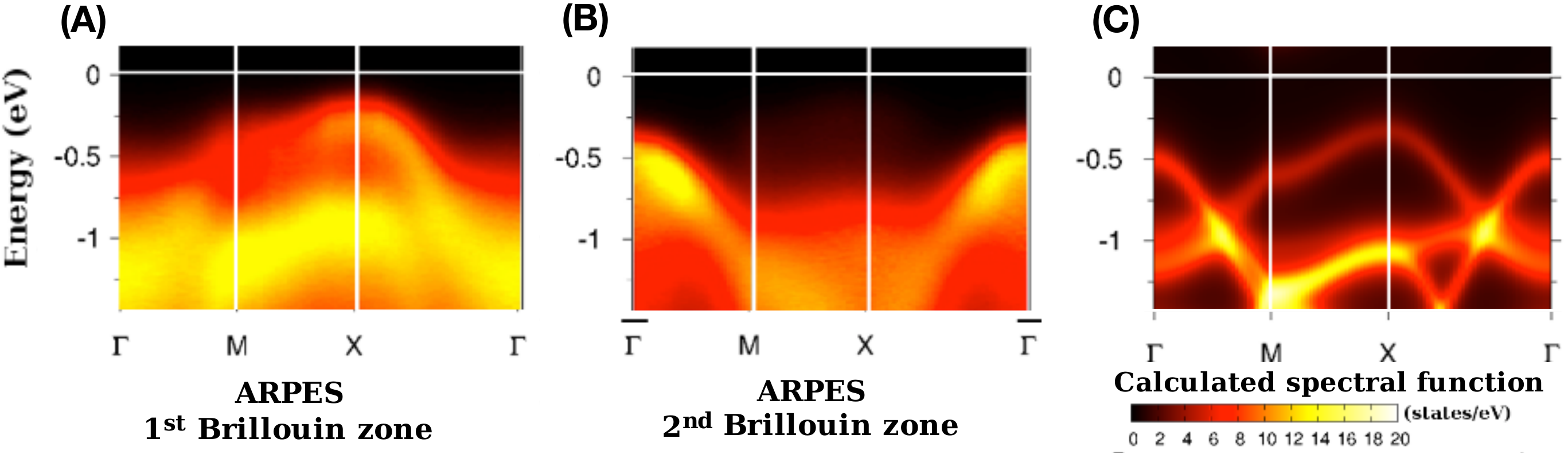}
	\caption{\label{Fig:PMSpec}ARPES spectra at room temperature in the first (A) and second (B) Brillouin zone, as well as the calculated spectrum (C), adapted from Martins et al. 2018 \cite{Martins2018}. Since the calculation does not include matrix-element effects, one should compare the weighted combination of spectra (A) and (B) with spectrum (C). Figure adapted from \textit{Martins et al.} \cite{Martins2018}.}
\end{figure}

Figure~\ref{Fig:PMSpec}(A) and (B) show the ARPES spectra taken at room temperature along the high-symmetry path $\Gamma-M-X-\Gamma$ for the first and second Brillouin zone, compare also with Fig.~\ref{Fig:Intro1}(B).
Although the rather high temperature causes considerable thermal broadening, different bands can still be identified.
In particular, the spectra in the first and second Brillouin zone differ drastically due to matrix element effects:
In the first BZ spectral weight of the \jeff12 band is enhanced whereas the second BZ is clearly dominated by contributions of the \j32 manifold.
Consequently, the first removal state at $\Gamma$, which is of \j32 character, is found in the second BZ and the first removal state at $X$ in the first BZ.

In Fig.~\ref{Fig:PMSpec}(C) the calculated spectrum in the PM phase is shown along the same path in the first BZ.
Due to the rotations of the oxygen octahedra the unit cell is doubled as compared to the undistorted structure (see, e.g., Ba${}_2$IrO${}_4$ \cite{Moser2014}) and the bands are backfolded at the boundary of the (reduced) Brillouin zone.
In the theory, by definition, there is no difference between first and second BZ.
A quantitative comparison would require a calculation of the photoemission matrix elements which is beyond the scope of the present work.
Nevertheless it is clear that a meaningful qualitative comparison can be obtained by assuming that features of the spectral function should show up in experiment either in the first or the second BZ.

Comparing with the calculated spectra, the spectral feature in the first BZ at the bonding energy $E_B\sim1$eV between $M$ and $X$ can be clearly attributed to be of \j32 character.
More precisely, the features of \j32 character in the first BZ stems from the $m_j=3/2$ band whereas the spectrum in the second BZ originates from the $m_j=1/2$ band.

The stark differences between spectra in the first and second BZ indicate that the symmetry breaking potential which doubles the unit cell is rather weak: 
In case of strong symmetry breaking that reduces the symmetry from $I_4/mmm$ to $I_4/acd$, the two Brillouin zones would show very similar spectra.
Here, however, the distortions of the oxygen octahedra introduce only a small Fourier component of the corresponding potential.
In addition, the polarization of the photons naturally leads to matrix element effects that enhance or suppress spectral weight in an orbital-selective manner \cite{Louat2018b,Louat2018}.

In contrast to single-site DMFT \cite{Martins2018} or LDA+U+SOC \cite{Wang2013} spectra, the present OC-DMFT spectrum correctly captures the absence of a weakly dispersive branch of \jeff12 character at the $\Gamma$ point with an onset at $-0.2$eV below the Fermi energy.
As can be seen from figure~\ref{Fig:PMSpec}(B) the onset of spectral weight at the $\Gamma$ point is only at $-0.5$eV and should have rather \j32 character showing a drastic difference to the spectrum calculated in DMFT.
The reason of this improvement lies in the explicit inclusion of inter-site fluctuations:
Antiferromagnetic short-range fluctuations within the dimer provide the previously missing ingredient.
The energy scale of this inter-site fluctuations can be estimated as described in Ref.~\cite{Martins2018} as $108$meV ($78$meV) for the interlayer (anti)bonding \jeff12 states.
This values compare well with an experimental estimate for the magnetic exchange coupling in the paramagnetic phase $J\sim 0.1$eV obtained from the temperature dependence of the in-plane correlation length within resonant magnetic x-ray scattering \cite{Fujiyama2012}.

%%%%%%%%%%%%%%%
\begin{figure}[h]
	\includegraphics[width = .5\linewidth]{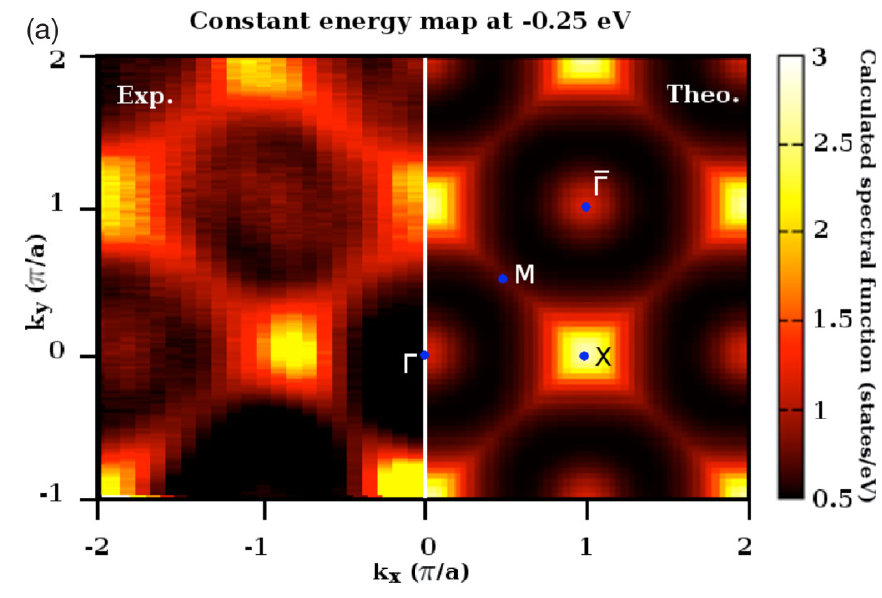}
	\includegraphics[width = .5\linewidth]{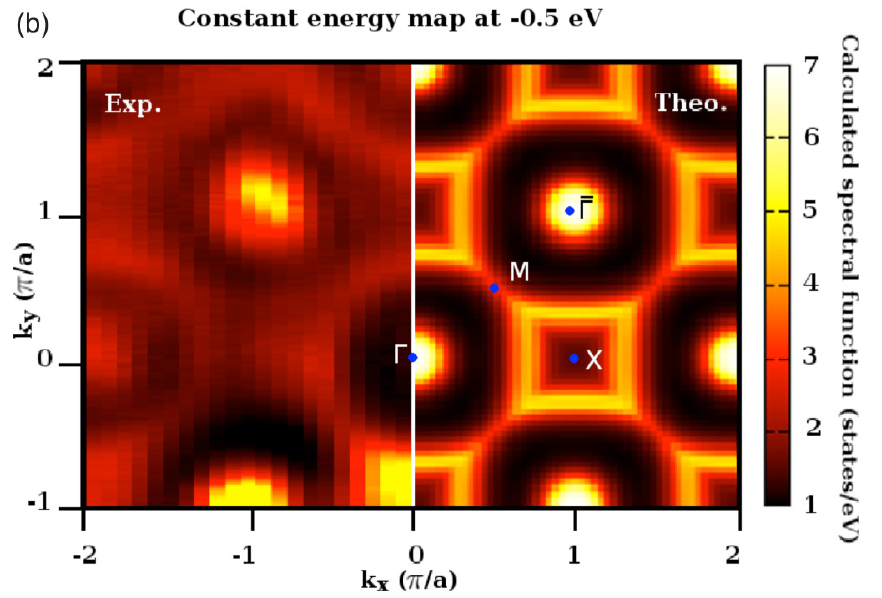}
	\caption{\label{Fig:PMEcut}Constant energy maps at binding energies $E_B=0.25$eV and $E_B=0.5$eV measured in experiment (left) and calculated within OC-DMFT (right), adapted from \textit{Martins et al.} 2018 \cite{Martins2018}. The high-symmetry points of the Brillouin zone are indicated by blue points. The first and second Brillouin zone are the same in the calculation, but differ due to matrix-element effects in experiment.}
\end{figure}
To compare the ARPES spectra further with calculated spectral functions, figure~\ref{Fig:PMEcut} shows equal energy maps at $E_B=0.25$eV and $E_B=0.5$eV.
The constant energy maps are very similar to those taken below $T_N$ in the antiferromagnetic phase \cite{Kim2008,Wang2013,Cao2016}, which shows that the photoemission spectra are not sensitive to long-range magnetic order.
 At $E=-0.25$eV, the peak at $X$ due to the \jeff12 band is well visible in both experiment and theory, whereas the weak precursor at the $\Gamma$ point is barely visible in the experiment and shows intensity modulations between different BZs.
For $E=-0.5$eV, the intense blobs at $\Gamma^{\prime}$ and much weaker peaks at $\Gamma$ show strong matrix element effects in the ARPES spectrum.
Apart from these matrix element effects, both spectra are again in good agreement.

%%%%%%%%%%%%%%%%%%%%%%%%%%%%%%%%%%%%%%%%%%%%%%%%%
\section{\sriro\ in the Antiferromagnetic Low-Temperature Phase}
\label{Sec:AF}
So far we discussed the paramagnetic phase of pure \sriro\ at temperatures above the N{\'e}el temperature of $T_N\sim230$K \cite{Kim2008}.
Below $T_N$ the \jeff12 pseudospins of \sriro\ order and a phase transition towards an antiferromagnetic phase occurs \cite{Kim2009}.
Since the moments are slightly tilted out of plane, the ground state is rather a canted antiferromagnet \cite{Ye2013,Dhital2013}.
In this section we will discuss the spectral properties of the antiferromagnetically ordered low-temperature phase of \sriro.

%%%%%%%%%%%%%%%
\subsection{Spectra from Theory}
\begin{figure}[h]
	\includegraphics[width = .9\linewidth]{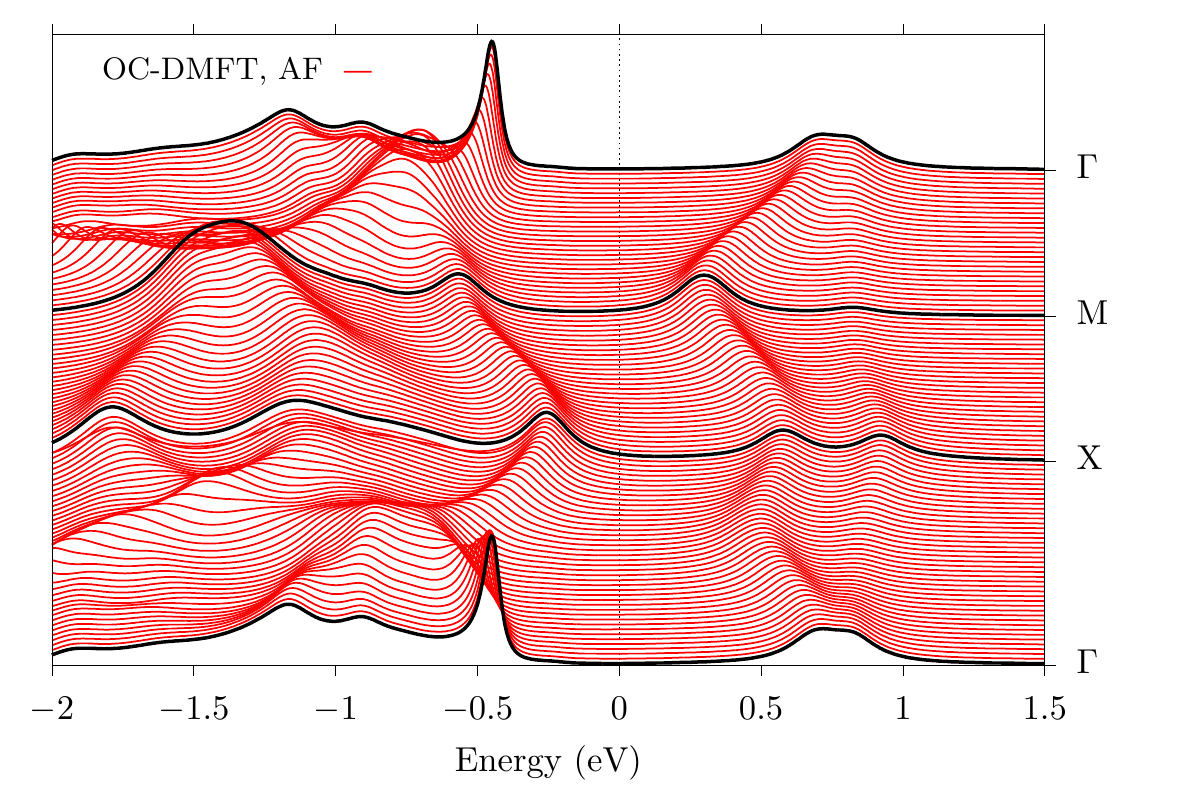}
	\caption{\label{Fig:AFTheo}Calculated spectral function $A(\mathbf{k},\omega)$ along the $\mathbf{k}$-path $\Gamma-M-X-\Gamma$ in the antiferromagnetic low-temperature phase.}
\end{figure}

Starting from the same effective one-band Hubbard model, but allowing for antiferromagnetic ordering within the dimers of the different planes leads in the OC-DMFT treatment to an AF ground state at low temperatures.
In our cluster treatment we assume an antiferromagnetic alignment of the $j_{\mathrm{eff}}=1/2$ moments on neighboring Ir spins.
This is done by including an antiferromagnetic Weiss field on the cluster, whose strength is determined via a variational principle using self-energy functional theory \cite{Potthoff2003,Dahnken2004}.
One should note that the experimentally measured magnetic configuration shows additional spin canting and leads to an effective weak ferromagnetic moment \cite{Ye2013}.
From a modeling point of view this can be included via a twisted Hubbard model \cite{Senthil2011}.
Since we focus here rather on spectral than on magnetic properties we keep an antiferromagnetic alignment of the moments, which -in principle- slightly overestimates the magnetization.
The corresponding spectral function for pure \sriro\ in this AF configuration is shown in figure~\ref{Fig:AFTheo}.

The antiferromagnetic spectrum is very similar to a paramagnetic spectrum calculated for the same temperature. 
This similarity is due to antiferromagnetic fluctuations, which are the essential ingredient for a precise description of the spectral function of paramagnetic \sriro.
Compared to the paramagnetic spectrum, the splitting between (anti)bonding band and (upper) lower Hubbard satellite band is a bit enhanced.
As a consequence, the two different contributions to the hole part of the spectrum at $\Gamma$ can now be identified, whereas the paramagnetic spectrum of Fig.~\ref{Fig:PMSpec_1}(C) shows only one single peak at $\sim0.6$eV.
Furthermore, some spectral weight is transferred from the (anti)bonding bands to the satellites.
Although it is only a minor change, the antiferromagnetic ordering leads thereby to additional spectral weight between $X$ and $M$ at $\sim-1$eV.

%%%%%%%%%%%%%%%
\subsection{Experimental Spectra}
\begin{figure}[h]
	\includegraphics[width = \linewidth]{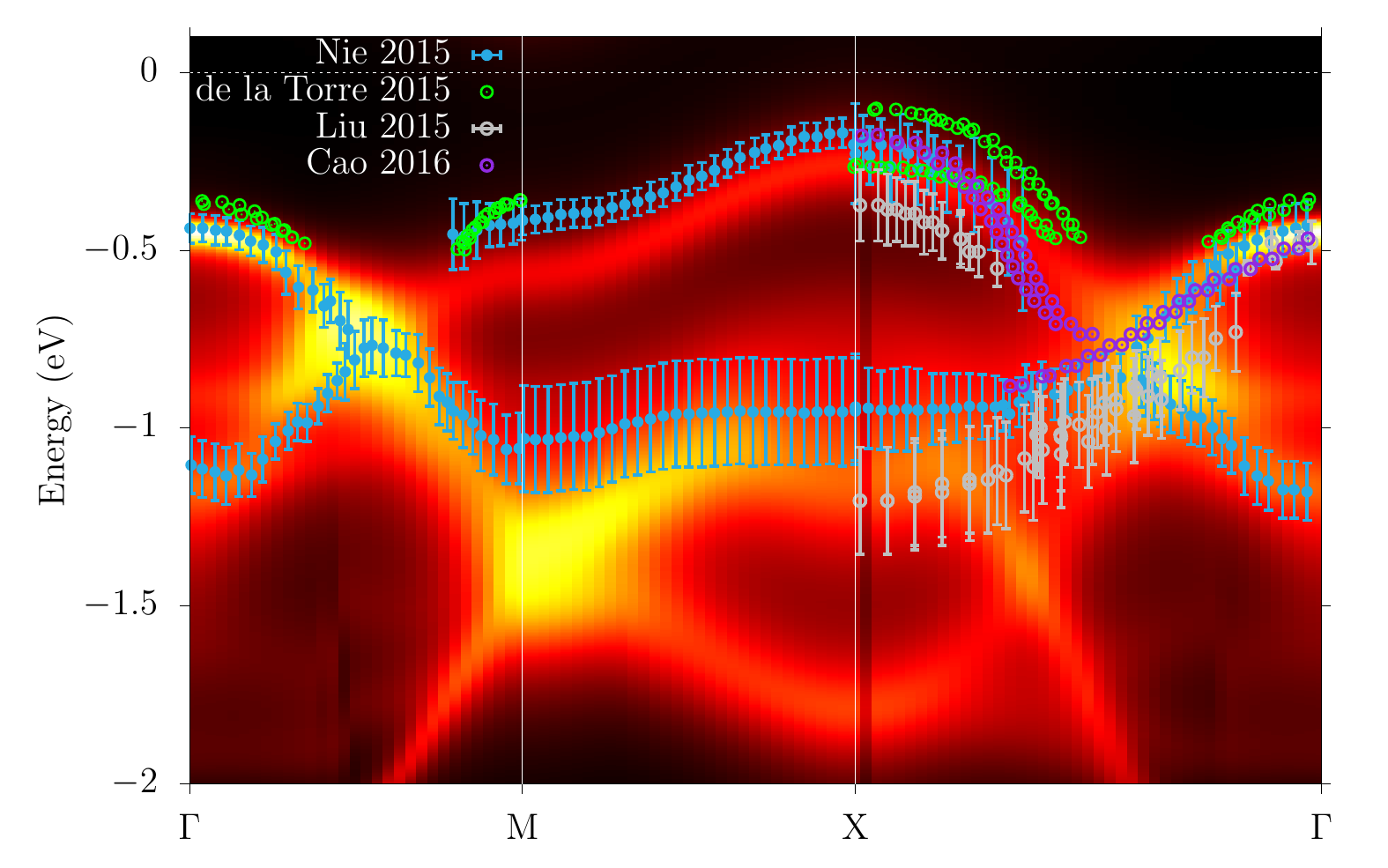}
	\caption{\label{Fig:AFExpTheo_spec}Comparison of the calculated spectral function $A(\mathbf{k},\omega)$ with dispersions extracted in ARPES works. Experimental data shown is extracted from \textit{Nie et al. 2015} \cite{Nie2015}, \textit{de la Torre et al. 2015} \cite{Torre2015}, \textit{Liu et al. 2015} \cite{Liu2015}, and \textit{Cao al. 2016} \cite{Cao2016}.}
\end{figure}

Using angle-resolved photoemission spectroscopy to investigate pure \sriro\ is limited to moderate temperatures due to surface charging effects that prevent an investigation of the sample at very small temperatures.
Nevertheless, spectra for temperatures down to $T=20$K have been measured \cite{Liu2015}, which is deep in the antiferromagnetic phase. 
Although the data quality is not as high as for doped samples, which are metallic and do not have this problem, the spectra allow for the extraction of dispersion that can be compared to our calculated spectrum, see Fig.~\ref{Fig:AFExpTheo_spec}.
Overall, the calculated spectrum is in good agreement with the different experimental data sets.
Since the experimental data shown were taken under different experimental conditions like temperature, photon energy or polarization, the dispersions vary a bit.
This is in particular visible at the upper band edge close to the X point, where the position of the \jeff12 band maximum varies by several $100$meV between the different works.

\begin{figure}[h]
	\includegraphics[width = .32\linewidth]{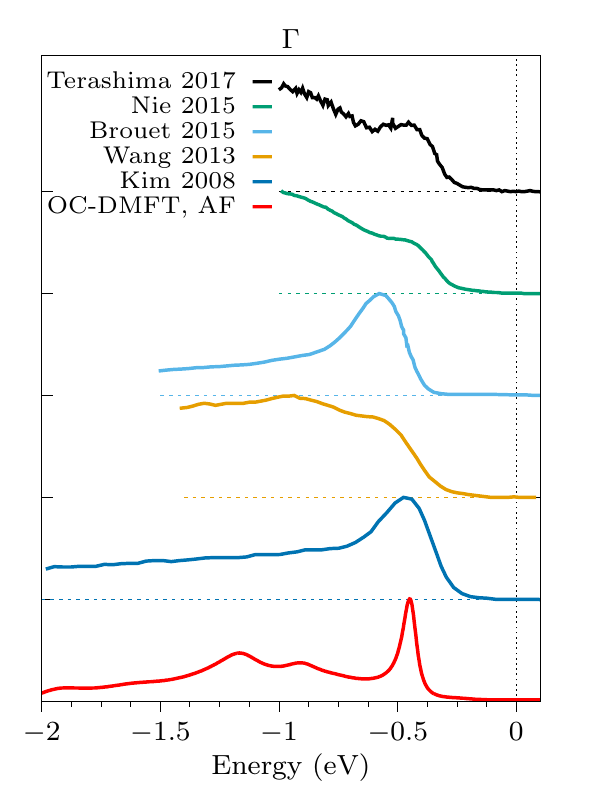}
	\includegraphics[width = .32\linewidth]{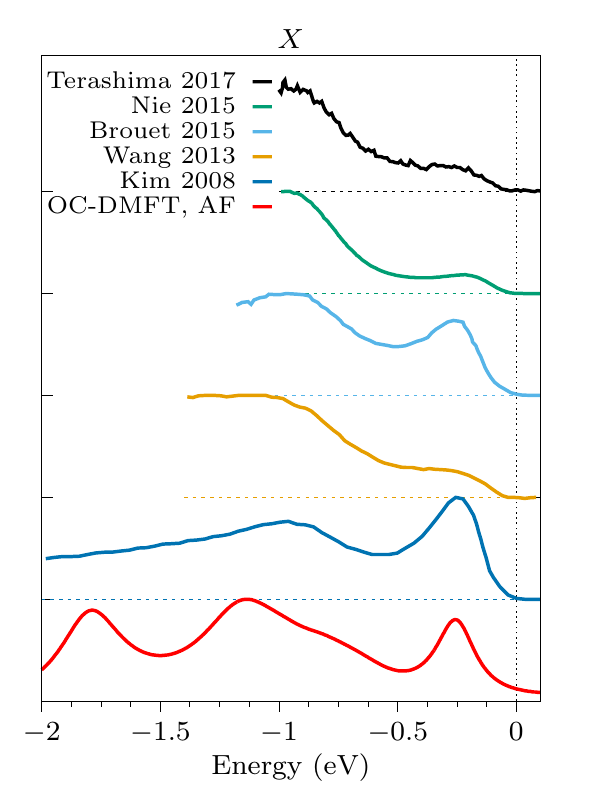}
	\includegraphics[width = .32\linewidth]{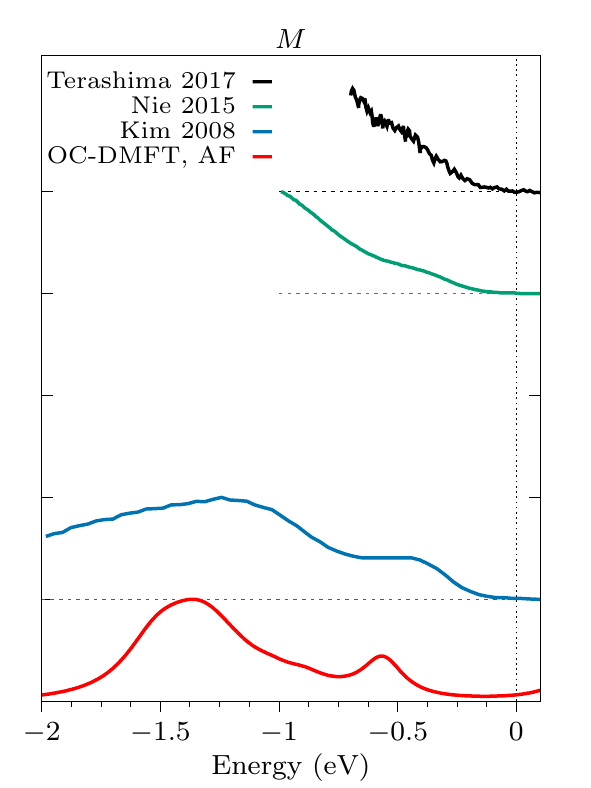}
	\caption{\label{Fig:AFExp_EDC}Detailed comparison of the calculated spectral function (red) and different ARPES measurements at the high-symmetry $\mathbf{k}$-points $\Gamma,X,$ and $M$. The spectral weight curves have been normalized with respect to the available maximal weight of the respective curves and dashed lines indicate the zero-intensity offsets of the respective curves. Some key properties of the experimental measuring conditions used in these works can be found in table \ref{Tab:ARPES_AF}. Experimental data are extracted from \textit{Kim et al. 2008} \cite{Kim2008}, \textit{Wang et al. 2013} \cite{Wang2013}, \textit{Brouet et al. 2015} \cite{Brouet2015}, \textit{Nie et al. 2015} \cite{Nie2015}, and \textit{Terashima et al. 2017} \cite{Terashima2017}.} 
\end{figure}

In order to compare the spectra more quantitatively, we plot the calculated spectral function at the high-symmetry points $\Gamma, M,$ and $X$ with the energy distribution curves (EDC) of different experimental studies in figure~\ref{Fig:AFExp_EDC}. 
Again, the calculated spectral function reproduces all features of the experimental EDCs at least qualitatively.
Modulations at larger bonding energies $E \leq -1$eV are more pronounced in the calculated spectra than in experiment, but the low-energy region compares favorably.
At $\Gamma$ and $X$ the onset of spectral weight and the position of maximal intensity agrees well between calculation and experiment.
However, at $M$ the peak position in the calculated spectrum is at smaller energy than in experiment.
One might think of two reasons for this deviation.
First, it is known that antiferromagnetism is overestimated within a dimer cluster and therefore leads here to a stronger splitting between bonding and antibonding band. 
Secondly, the ground state of \sriro\ at low temperatures is a canted antiferromagnet whereas on the cluster level the system adopts a purely antiferromagnetic configuration with larger net staggered magnetization.

Quite generally one observes that the spectra in PM and AF phases are very similar \footnote{Interestingly, a similar trend has also been found in the undoped cuprate Sr${}_2$CuO${}_2$Cl${}_2$, where ARPES spectra in PM and AF phases are much akin \cite{Wallauer2015}.}.
This is consistent with our finding that our dimer approach, which introduces AF fluctuations, provides an extremely good description of the PM. 
It also explains the success of AF DMFT calculations even in its single-site version to describe both phases.
Note that an AF ordered dimer is well described by a mean-field description and the AF ordered insulator can -in this sense- be considered a band insulator.
This effect explains the apparent success of DFT+U-like descriptions of spectra of \sriro\ \cite{Kim2008,Zhang2013}, which however strongly differ from experiment at the $\Gamma$ and M point.

%%%%%%%%%%%%%%%
\subsection{Constant Energy Cuts}
\begin{figure}[h]
\begin{center}
	\includegraphics[width = \linewidth]{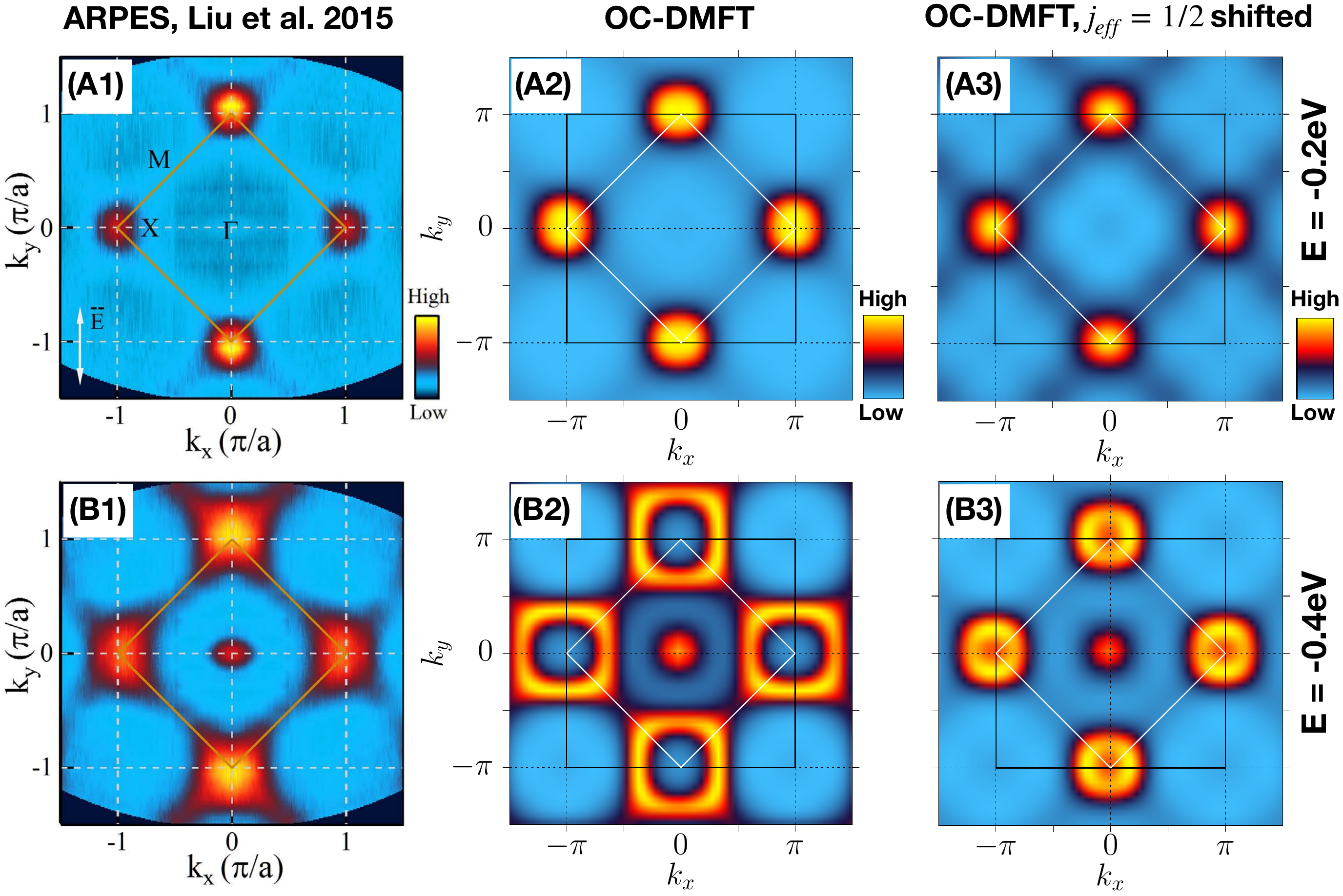}
\end{center}
	\caption{\label{Fig:AFExp_Ecuts} {Constant energy maps of \sriro\ in the antiferromagnetic phase at energies $E=-0.2$eV (A) and $E=-0.4$eV (B). The angle-resolved photoemission spectra at $T\sim20$K (A1,B1) are adapted from \textit{Liu et al.} 2015 \cite{Liu2015} and are compared to OC-DMFT spectra (A2,B2) at the same binding energies. (A3,B3) show the calculated spectra with an additional shift of $-0.15$eV for the \jeff12 manifold, see text.}}
\end{figure}

In figure \ref{Fig:AFExp_Ecuts} we compare the calculated constant energy maps of \sriro\ in the antiferromagnetic phase at two different binding energies with ARPES spectra from Ref.~\cite{Liu2015}.
Qualitatively, the features are well captured by the OC-DMFT calculation: At $E=-200$meV only the \jeff12 bands at the X points have non-zero spectral weight and theory and experiment only differ in the $x-y$ anisotropy of spectral intensity in experiment, which stems from matrix-element effects that are not taken into account in the calculation.
Due to linear polarization of the photon beam in $k_y$ direction the experimental spectra have larger spectral weight around X-points in $k_y$- than in $k_x$-direction, whereas the calculated spectra still show $x-y$ symmetry.
Compared to Fig.~\ref{Fig:AFExp_Ecuts}A, at a binding energy of $E=400$meV both experiment and theory show in addition some spectral weight at the $\Gamma$ point due to the \j32 band, which is peaked at slightly larger binding energies.
The main contribution of spectral weight still stems from the \jeff12 band, however, in contrast to the measurement the calculated spectrum is not peaked at the X points, but rather around them.
Good agreement with the \jeff12 part of the measured spectrum would rather be obtained by assuming that the \jeff12 part of the experimental spectrum corresponds in fact to slightly smaller binding energies than indicated.
This can be seen in the third column of Fig.~\ref{Fig:AFExp_Ecuts}, where the \jeff12 part of the calculated spectrum is in addition shifted by $-0.15$eV. 

This is consistent with the comparison to other ARPES works shown in Fig.~\ref{Fig:AFExpTheo_spec}, where one sees that the dispersion of Ref.~\cite{Liu2015} is located at the X point at $\sim0.1$eV larger binding energies.
We note that the dispersion at the $\Gamma$ point is not affected, but rather consistent between the different ARPES data shown in Fig.~\ref{Fig:AFExpTheo_spec}. 
Therefore, the origin of the apparent slight difference of the energetic difference between \jeff12 and \j32 bands in the spectra of \textit{Liu et al.} as compared to theory remains a puzzle.

%%%%%%%%%%%%%%%%%%%%%%%%%%%%%%%%%%%%%%%%%%%%%%%%%
\section{Electron-doped \sriro\ at Low Temperatures}
\label{Sec:ElDop}
The similarity of its electronic structure with hole-doped high-temperature cuprate superconductors has sparked interest in electron-doped \sriro\ and triggered experimental and theoretical investigations in recent years.
In contrast to the cuprates, no superconductivity has been found in \sriro, it rather becomes a paramagnetic metal down to lowest temperatures for sufficiently strong doping. 
Nevertheless, spectral properties like pseudogap features inherent to cuprates have also been proposed for doped \sriro\ \cite{Senthil2011,Torre2015}.
In this section, we will discuss the spectra of electron-doped \sriro\ at low temperatures calculated from theory and compare it to ARPES measurements of La-doped samples.

%%%%%%%%%%%%%%%
\begin{figure}[h!]
	\includegraphics[width=\linewidth]{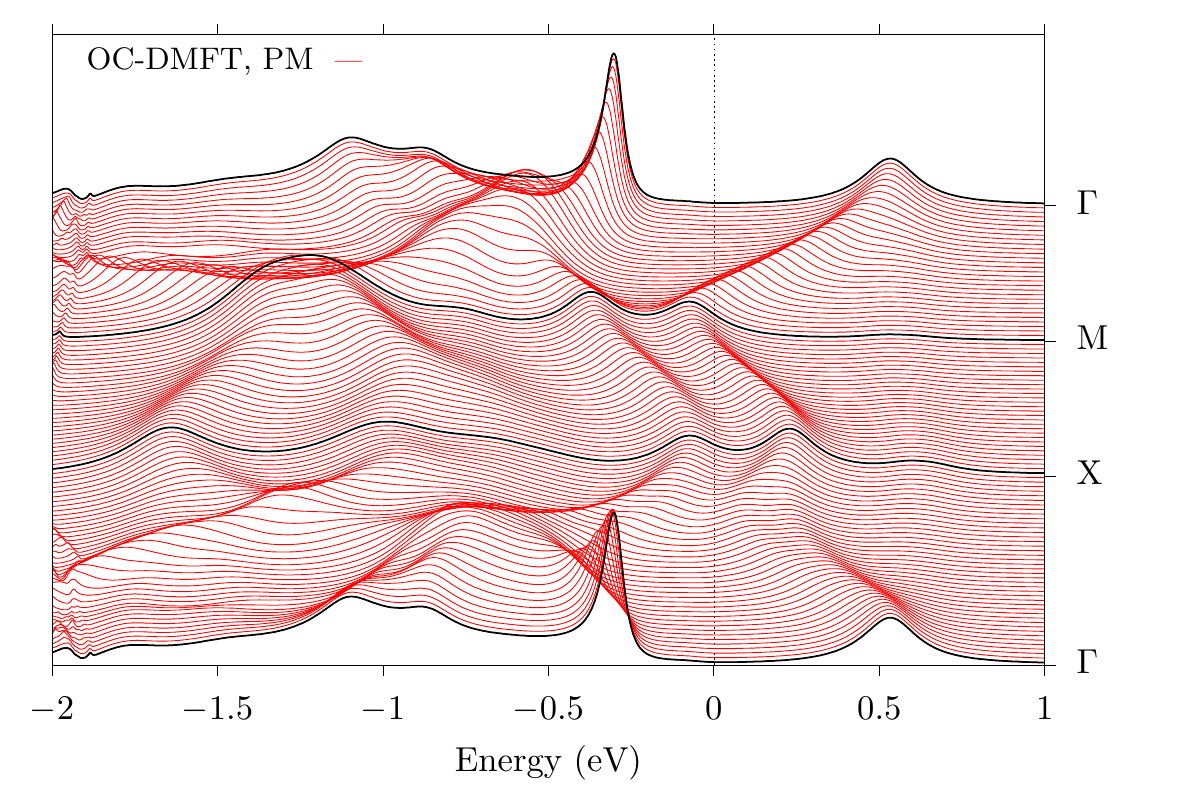}
\caption{\label{Fig:SpecElDop}Spectral function $A(\mathbf{k},\omega)$ of $10\%$ electron-doped \sriro along the $\mathbf{k}-$path $\Gamma-M-X-\Gamma$. Within this paramagnetic calculation we used an effective Coulomb interaction strength of $U=0.6$eV.}
\end{figure}

\subsection{Theoretical Spectra}

In order to calculate the spectral function of electron-doped \sriro\ we apply OC-DMFT to the \jeff12 band at an electron filling of $n=0.9$.
It should be noted that the \j32 bands are still completely filled and therefore not much affected by the change in the electron filling $n$, see Fig.~\ref{Fig:SpecElDop}.

For the \jeff12 band the change in electron filling renders the system metallic.
In a metal, electronic screening is much enhanced as compared to an insulator and as a consequence, the effective local Coulomb interaction strength is reduced \cite{Ayral2013}.
Here, we take the enhanced screening into account by using an effectively reduced Coulomb interaction strength of $U_{\mathrm{eff}}=0.6$eV, which leads to good agreement of the calculated spectra with ARPES data.
The corresponding spectral function $A(\mathbf{k},\omega)$ is plotted in Fig.~\ref{Fig:SpecElDop}.
Due to the electron doping, the Fermi surface now has pockets around the M point and the lower \jeff12 branch is close to $E_F$ at $M$.
Note that we have not recalculated the position of the \j32 states in the doped case from DMFT.
Instead, we estimate an electrostatic shift of $\Delta E_{\mathrm{Hartree}}=U\Delta n$, which was applied to the bands calculated in the undoped case.

%%%%%%%%%%%%%%%
\begin{figure}[h]
	\includegraphics[width=\linewidth]{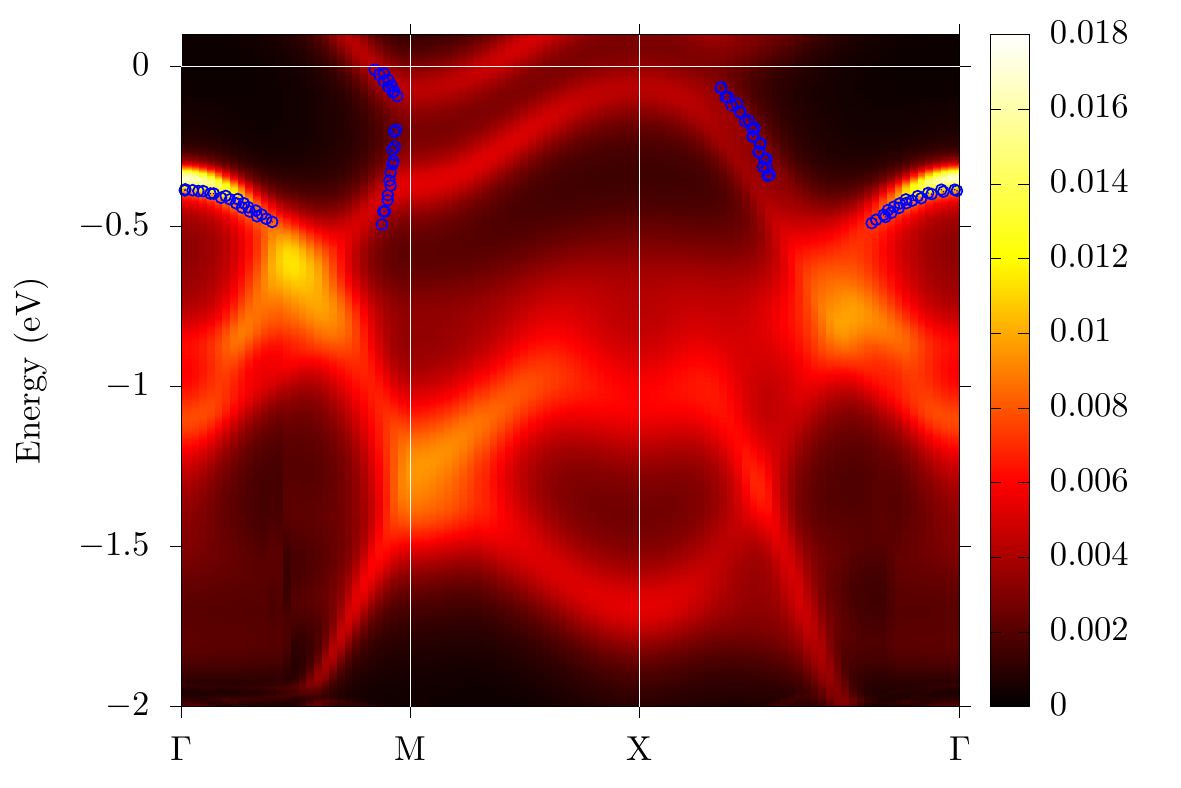}
\caption{\label{Fig:ExpSpec}Comparison of the calculated spectral function of $10\%$ electron-doped \sriro with ARPES data of a $x=0.05$ La-doped sample $(\mathrm{Sr}_{1-x}\mathrm{La}_x)_2$IrO${}_4$. Experimental data points (blue) are extracted from \textit{de la Torre et al.} \cite{Torre2015}.}
\end{figure}

\subsection{Experimental Spectra}

Most experimental works on electron-doped \sriro\ focus on the low-energy region at and close to the Fermi level $E_F$.
Before discussing two spectral features close to $E_F$ in more detail in the following two subsections, we compare in Fig.~\ref{Fig:ExpSpec} the calculated spectrum to available ARPES data of La-doped samples (Sr${}_{1-x}$La${}_x$)${}_2$IrO${}_4$.\footnote{Note that $x=0.05$ thereby amounts to 10\% electron doping.}

The spectral feature of \j32 character at the $\Gamma$ point is correctly captured, as well as the dispersion of the upper \jeff12 branch around the $M$ point, which leads to pocket structures in the Fermi surface.
Along the $\Gamma-X$ direction the \jeff12 dispersion is captured with small deviations from the ARPES data of Ref.~\cite{Torre2015}.
However, the calculated spectrum of the lower \jeff12 branch deviates close to $M$ from the measured photoemission spectrum and is discussed in the next subsection.

%%%%%%
\subsection{Spectra close to the M point}
\begin{figure}[h]
\begin{center}
	\includegraphics[width=.9\linewidth]{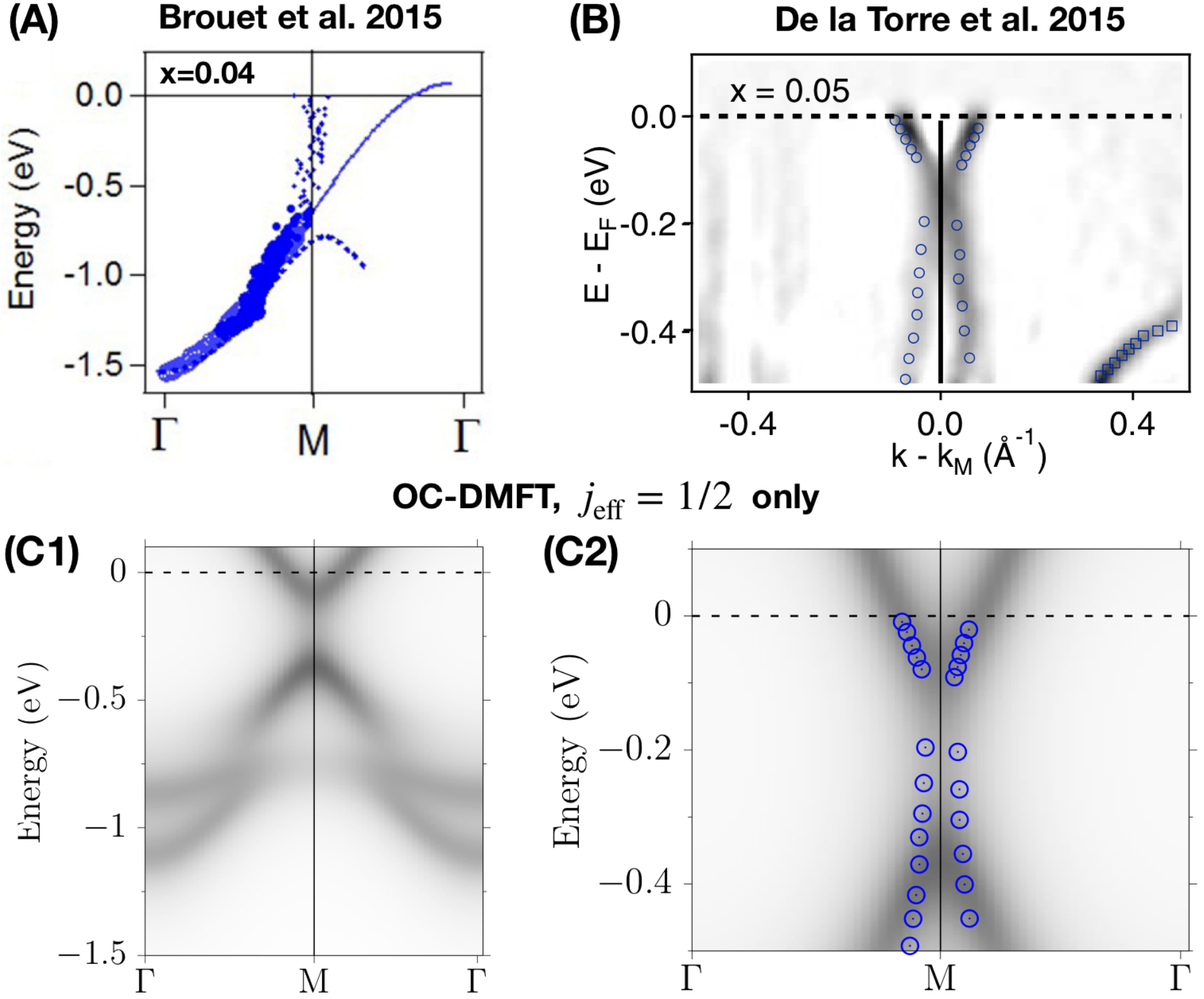}
\caption{\label{Fig:Doped_Mpt}Comparison of ARPES spectra (A,B) with calculated spectra (C1,C2) around the $M$-point. (A) The dispersion extracted by MDC analysis along $\Gamma$-M for a sample with $x-0.04$ La-doping, adopted from \textit{Brouet et al.} \cite{Brouet2015}. The dots represent peak positions extracted from Lorentzian fits and follow the solid blue line up to $-0.6$eV. Close to the M point the correlation gap is indicated by the dashed blue line. (B) Curvature plots of ARPES raw data for a $x=0.05$ doped sample, adapted from \textit{de la Torre et al.} \cite{Torre2015}. Blue dots indicate peak positions determined from this plot. (C1,2) show the calculated \jeff12 spectra for $10\%$ electron doped \sriro\ in the two energy ranges of (A) and (B). }
\end{center}
\end{figure}

In this subsection, we focus on spectral properties around the $M$ point, see Fig.~\ref{Fig:Doped_Mpt}.

In Ref.~\cite{Torre2015} \textit{de la Torre et al.} suggested that the bands with nearly linear dispersion close to the $M$ point  might extrapolate to a Dirac point at $E\sim-0.1$eV before they continue still with a quasi-linear dispersion up to the Fermi energy (Fig.~\ref{Fig:Doped_Mpt}(B)).
For a slightly less La-doped sample, \textit{Brouet et al.} found a very similar quasi-linear feature close to the $M$ point \cite{Brouet2015}, which is much reduced in intensity, but non-negligible.
However, they interpret this feature as an artifact of the fitting procedure applied to the momentum distribution curves (MDC):
Since the \jeff12 energy distribution curve has a quite broad peak at lower energy the tail of this peak can lead to a signal with quasi vertical dispersion that is picked up in the MDC analysis.
For the case of larger La-doping, where the states close to $E_F$ clearly form a pocket around $M$, this suggests also a different interpretation of the signal for binding energies $E_B>0.1$eV.
Instead of a Dirac point of two connected bands, the region below $E\sim-0.1$eV could be gapped and the spectral weight just be picked up by the broad tail of the lower \jeff12 band at lower energy.

To investigate this question we plot the \jeff12 part of the calculated spectral function for $10\%$ electron doping in Fig.~\ref{Fig:Doped_Mpt}(C1,2) once in a large energy region and once in close vicinity of the Fermi energy.
As can be seen from (C1) the OC-DMFT spectrum is gapped in the region $-0.35\mathrm{eV}\leq E\leq-0.1\mathrm{eV}$ and has large spectral weight at $M$ and $E\sim-0.4$eV.
Along roughly half the distance $M-\Gamma$ the spectral features are in good agreement with (A), close to $\Gamma$ spectral contributions from \j32 render the comparison more difficult, but main features are qualitatively still comparable to (A).
 When comparing to (B) one notices good agreement for the band forming the pocket in the immediate vicinity of the Fermi energy.
 Also for energies $E\leq-0.35$eV the spectrum agrees well with (B) suggesting a reinterpretation of the ''gapped'' region $-0.35\mathrm{eV}\leq E\leq-0.1\mathrm{eV}$ in terms of Ref.~\cite{Brouet2015}.
 
%%%%%%%%%%%%%%%
\subsection{Fermi Surface, Electron Pockets and Pseudogap}
\begin{figure}[h]
	\includegraphics[width=\linewidth]{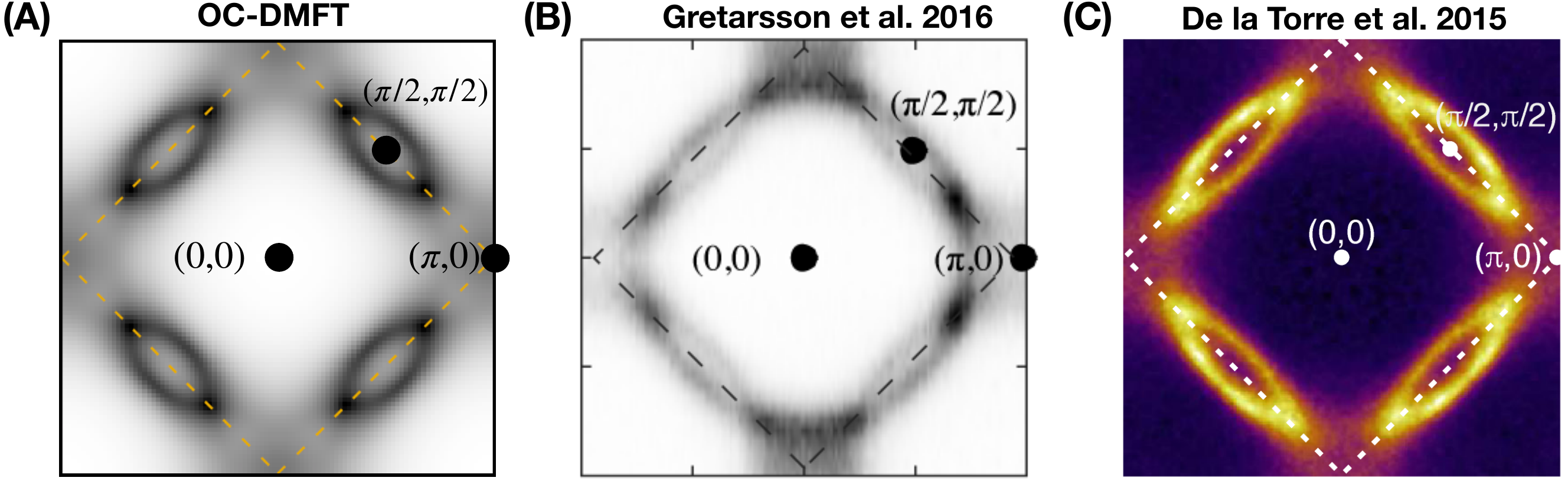}
\caption{\label{Fig:Pockets}Fermi surface of $5\%$ La-doped \sriro\ within ARPES showing the Fermi pockets at $T=10$K (B), adopted from \textit{Gretarsson et al.} \cite{Gretarsson2016}, and at $T=8$K (C), adopted from \textit{de la Torre et al.} \cite{Torre2015}. The calculated Fermi surface at $10\%$ electron-doped \sriro\ is shown in (A).}
\end{figure}

For La-doping $x\geq0.05$ the system is a paramagnetic metal with a Fermi surface with well-defined electron pockets.
In Fig.~\ref{Fig:Pockets} we compare ARPES spectra at the Fermi energy of $5\%$ La-doped \sriro\ with the calculated Fermi surface at $10\%$ electron doping.
The electron pockets around the $M$ points are lens-shaped with their elongated axis pointing towards the $X$ points.
Comparing position, shape and size of the pockets in (A) with the experimental observations leads to overall good agreement. 
Most spectral weight of the pockets is found at their tips, consistent with experiment ((B) and (C)).
By symmetry, both sides of the pockets carry equal spectral weight in the theoretical spectral function.
In the ARPES spectra, the spectral weight in the second BZ is enhanced as compared to the backfolded part of the pocket.

\begin{figure}[h]
\begin{center}
	\includegraphics[width=.9\linewidth]{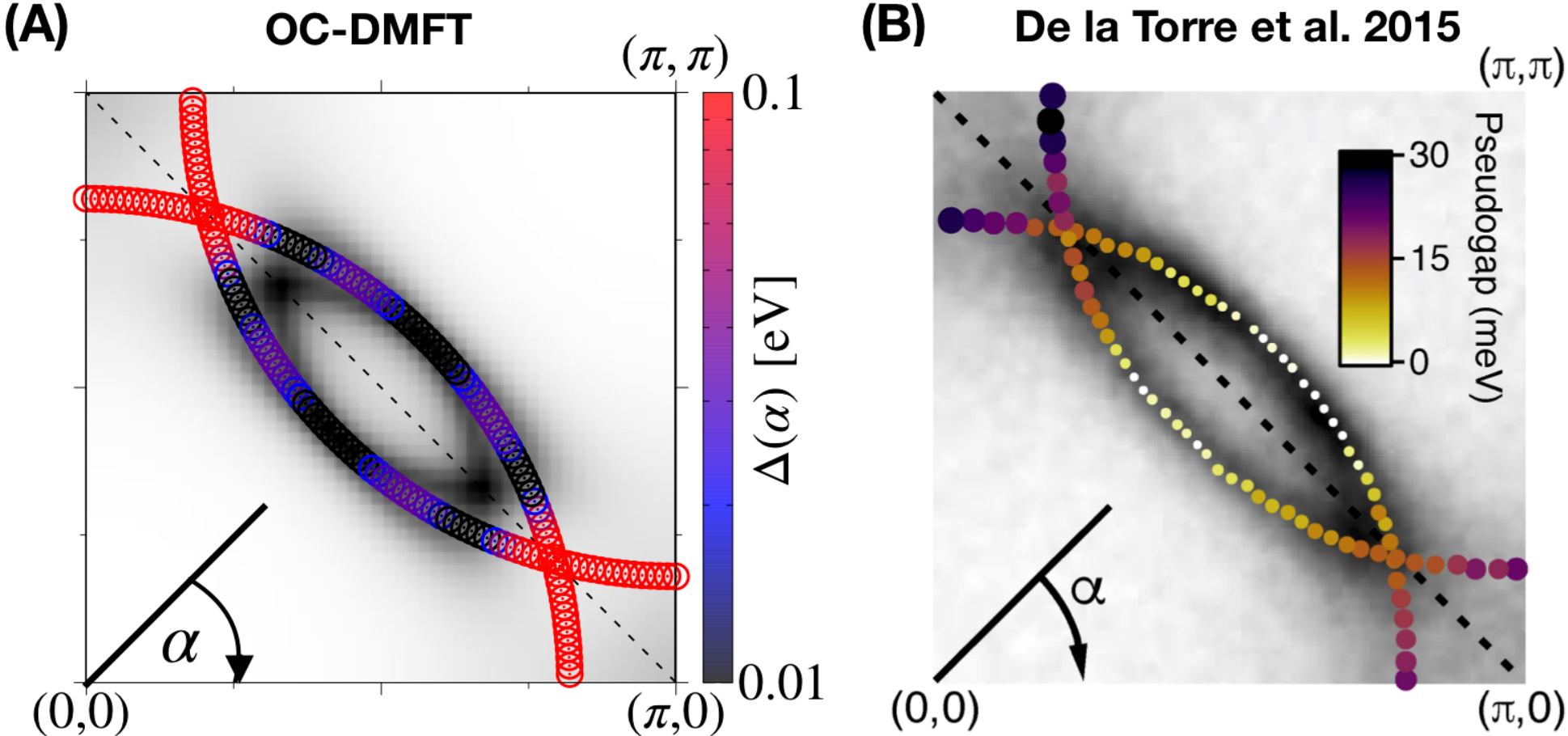}
\caption{\label{Fig:PG}Calculated $\Delta(\alpha)$ of $10\%$ electron-doped \sriro\ (A) and measured pseudogap (B) of a $5\%$ La-doped sample at $T=8$K. In both cases, the quantities were extracted along the remnant Fermi surface as discussed in the text. Panel (b) is adopted from \textit{de la Torre et al.} \cite{Torre2015}, BF denotes back-folded.}
\end{center}
\end{figure}

In  experiments, resistivity measurements of La-doped samples display unusual metallic behavior at low temperatures \cite{Brouet2015,Torre2015} and for $x=0.05$ doped samples an anti-nodal pseudogap was found using ARPES \cite{Torre2015}, shown in Fig.~\ref{Fig:PG}(B). 
Since in ARPES the hole-part of the spectrum cannot be accessed, one often uses the particle-hole symmetrized spectral function to extract pseudogap features. 
To compare with experimental findings, we plot this symmetrised spectral function in Fig.~\ref{Fig:PG}.
Along the remnant Fermi surface, which in this case is a nearly circular path in k-space, one then extracts the ''gap'' $\Delta(\alpha)$ of the symmetrised spectral function as a function of angle $\alpha$.
This gap is defined by the peak position $\omega^{*}$ of this spectral function closest to the Fermi level at a given $\mathbf{k}$-point, $\Delta(\alpha) = 2\omega^{*}(\mathbf{k}_{\alpha})$.

Comparing the calculated $\Delta(\alpha)$ in Fig.~\ref{Fig:PG}A with the pseudogap (PG) extracted in Ref.~\cite{Torre2015} from ARPES measurements (Fig.~\ref{Fig:PG}B) shows qualitatively similar features.
Both $\Delta(\alpha)$ and the PG emerge at the tips and outside of the elliptical pocket along the circular $\mathbf{k}$-path.
In addition, within the calculations even the pocket structure itself at small angles $\alpha$ shows regions of non-zero $\Delta(\omega)$.
These spectral features can be tracked back to the average over the two perpendicular cluster orientations used within the OC-DMFT calculation \cite{Martins2018}.
To observe them separately would, however, require a very high resolution in experiment.
Furthermore, the origin of the large $\Delta(\omega)$ values close to the antinodal point can be identified in the calculation by comparing to the full spectral function including the hole part.
In contrast to the pseudo-gap-like features close to the tips of the pockets, the antinodal point shows a true gap since the spectral peaks closest to the Fermi energy belong to the \j32 band.

%%%%%%%%%%%%%%%%%%%%%%%%%%%%%%%%%%%%%%%%%%%%%%%%%
\section{Conclusion and Outlook}
\label {Sec:Out}
In this work we gave an overview of cluster-DMFT-based calculations of spectral functions of pure and electron-doped \sriro.
For undoped \sriro\ we obtained good agreement with ARPES spectra both in the antiferromagnetic and paramagnetic phase, and in the electron-doped system pocket structures of the Fermi surface and prominent pseudogap features of the spectrum could be reproduced.

As most theoretical works on \sriro, also our work here is based on effective one-orbital descriptions of the correlated \jeff12 bands and lead to good agreement with experiment.
However, both \textit{ab initio} calculations \cite{Martins2010} and angle-resolved photoemission spectra \cite{Louat2018,Louat2018b} revealed a strong $\mathbf{k}$-dependence of the $t_{2g}$ contributions that form the effective \jeff12 band.
One of the challenges to future studies is to extend the current description to polarization-dependent ARPES spectra, where effects not captured in a local \jeff12 picture could enter into play.

Furthermore, a question that has been raised recently is whether hidden orders such as the time-reversal symmetry breaking hidden order found in neutron scattering experiments \cite{Jeong2017} or the (different) hidden order investigated in Ref.~\cite{Zhou2017} might influence spectral properties.

%%%%%%%%%%%%%%%%
\ack
This work was supported by a Consolidator Grant of the European Research Council (Project CorrelMat-617196), the French Agence Nationale de la Recherche under project SOCRATE (Project No. ANR-15-CE30-0009-01) and  supercomputing  time at IDRIS/GENCI Orsay (Project No.  t2018091393).
We acknowledge helpful discussions with Steffen Backes, Fabrice Bert, Dalila Bounoua, Philippe Bourges, Veronique Brouet, Jaehong Jeong, Alex Louat, Luca Perfetti, and Yvan Sidis and we are grateful to the CPHT computer support team.

%%%%%%%%%%%%%%%%

\appendix

\section{Tight-binding model}
\label{App:TB}
To capture the physics of the DFT bands of Sr${}_2$IrO${}_4$, usual tight-binding models \cite{Jin2009,Carter2013,Moutenet2018} describe the $t_{2g}$ bands and then add spin-orbit coupling. 
In this work, we realized directly a tight-binding model for the $j_{\mathrm{eff}}$ bands to describe the DFT band structure, based on our understanding from the Wannier-projected characters (see Fig. 1).

Because of the distortions, there are 4 iridium atoms in the unit cell.
 In the following, we therefore consider a unit cell with two Ir dimers stacked along the $c$-axis.
 
Assuming no hopping between the $j_{eff}=1/2$ and $j_{eff}=3/2$ orbitals, one can restrict the description to the following $4\times 4$ matrix for the $j_{eff}=1/2$ bands:
\begin{eqnarray}
 H_0(\mathbf{k}) & = \left( \begin{array}{cc}
  \mathcal{O}_1(\mathbf{k}) & \mathcal{T}(\mathbf{k}) \\
  \mathcal{T}(\mathbf{k})^\dagger & \mathcal{O}_2(\mathbf{k}) 
  \end{array}\right).
\end{eqnarray}
Here, $\mathbf{k}$ is  expressed in terms of the reciprocal vectors forming the reduced Brillouin zone. 
The submatrix $\mathcal{T}(\mathbf{k})$ describes the inter-layer hopping part of the Hamiltonian:
\begin{equation}
 \mathcal{T}(\mathbf{k}) = -t_3 \left( \begin{array}{cc}
  \gamma_{k_y}+e^{-ik_z}\gamma_{k_x}^* & \gamma_{k_x}^*+e^{-ik_z}\gamma_{k_y} \\
  \gamma_{k_x}+e^{-ik_z}\gamma_{k_y}^* & \gamma_{k_y}^*+e^{-ik_z}\gamma_{k_x}
  \end{array}\right)
\end{equation}
with $\gamma_{k_x}=1+e^{+ik_x}$ and $\gamma_{k_y}=1+e^{+ik_y}$.
The submatrix $\mathcal{O}(\mathbf{k})$ describes the inter-layer part of the Hamiltonian, which is identical for the two layers:
\begin{eqnarray}
 \mathcal{O}_1(\mathbf{k}) = \mathcal{O}_2(\mathbf{k}) & = \left( \begin{array}{cc}
  e_\mathbf{k}^A & t_\mathbf{k} \\
  t_\mathbf{k}^{*} & e_\mathbf{k}^{B} 
  \end{array}\right)
\end{eqnarray}
where $e_\mathbf{k}^ A  = e_\mathbf{k}^B = \varepsilon_0-2t_2(\cos(k_x)+\cos(k_y)) + \varepsilon_{hyb}(\mathbf{k})$ and $t_\mathbf{k} = -2t_1e^{i\frac{k_x+k_y}{2}}\gamma_1(\mathbf{k})$ with $\gamma_1(\mathbf{k})=\cos(\frac{k_x+k_y}{2}) + \cos(\frac{k_x-k_y}{2})$. 
The additional term $\varepsilon_{hyb}(\mathbf{k}) = +4\varepsilon_1 \gamma_1(\mathbf{k})^2$ comes from the hybridization between the
former $d_{xy}$ and $d_{x^2-y^2}$ orbitals \cite{Jin2009} or can also be understood as an additional in-plane next-next-nearest neighbor hopping.

\begin{figure}[h]
\begin{center}
\includegraphics[width=.8\linewidth]{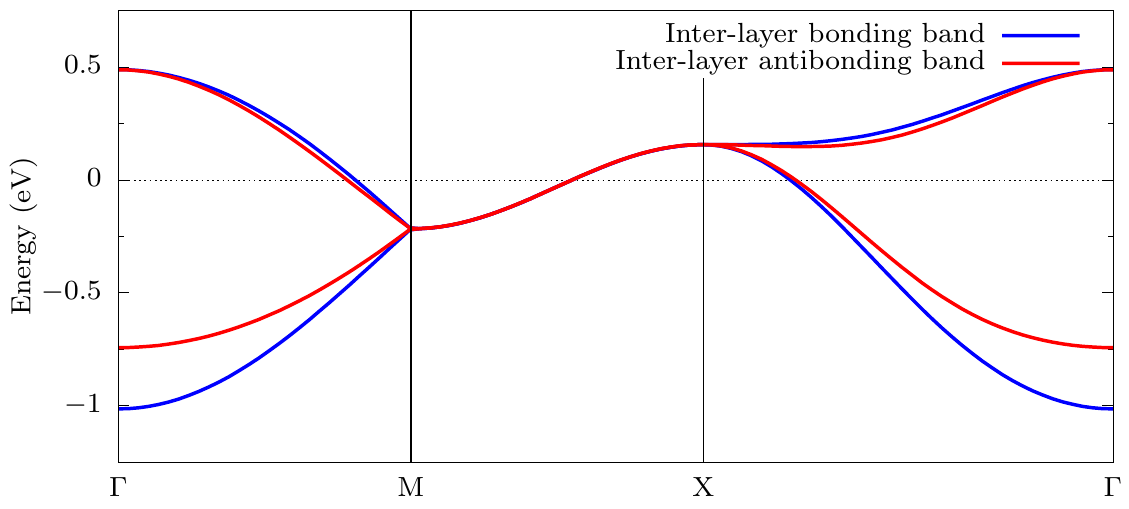}
\end{center}
\caption{\label{Fig:Inter-layer}Tight-binding fit of the DFT bands for the two-sheet unit cell consisting of two dimers per sheet, leading to an inter-layer bonding and inter-layer antibonding band in $c$-direction.}
\end{figure}

The best fit was obtained with $\varepsilon_0 = -0.237$~eV,$t_1 = 0.171$~eV, $t_2 = 0.0935$~eV, $t_3= 0.0169$~eV and $\varepsilon_1 =  0.026$~eV.
Due to the inter-layer hopping $t_3$, one obtains two bands: An inter-layer bonding and an inter-layer antibonding band.\footnote{Note that the inter-layer bonding-/antibonding bands should not be confused with the bonding-/antibonding- splitting within the dimer, which we refer to in the main text.} 
One can also downfold this effect and consider each layer independently. 
In this case, for the inter-layer bonding band, one gets the following parameters: $\varepsilon_0 = -0.217$~eV,$t_1 = 0.188$~eV, $t_2 = 0.0935$~eV and $\varepsilon_1 =  0.021$~eV , while for the antibonding band, one gets $\varepsilon_0 = -0.217$~eV,$t_1 = 0.154$~eV, $t_2 = 0.0935$~eV and $\varepsilon_1 =  0.029$~eV, see Fig.~\ref{Fig:Inter-layer}.

Just as in DFT calculations, the filling of these tight-binding dispersions is slightly larger than half-filling.
However, three-orbital DMFT calculations showed that treating the full \jeff12 and \j32 manifold leads to a charge-transfer into the \j32 bands and results in a half-filled \jeff12 band \cite{Martins2017}.
To set our one-band effective model, we take into account this charge-transfer by adjusting the chemical potential $\mu$ such that the \jeff12 manifold is half-filled. 

\section{Comparison of different ARPES studies on \sriro}
In Table \ref{Tab:ARPES_AF} we list the temperatures and photon energies at which the experimental ARPES measurements that were used in this article have been performed.
\begin{table}[h]
\begin{center}
\begin{tabular}{|c|c|c|c|}
\hline&Temperature $T$ (K)& Energy $h\nu$ (eV)\\\hline
\textit{Kim et al.} 2008 \cite{Kim2008} & 100 & 85\\
\textit{Wang et al.} 2013 \cite{Wang2013} & 25 & 80\\
\textit{Liu et al.} 2015 \cite{Liu2015} & 20 & 21.2\\
\textit{de la Torre et al.} 2015 \cite{Torre2015} & 50 & 100\\
\textit{Brouet et al.} 2015 \cite{Brouet2015} & 50 & 100\\
\textit{Nie et al.} 2015 \cite{Nie2015} & 70 & 21.2\\
\textit{Cao et al.} 2016 \cite{Cao2016} & 50 & 77,80,90\\
\textit{Terashima et al.} 2017 \cite{Terashima2017} & 100 & 100\\
\textit{Martins et al.} 2018 \cite{Martins2018} & 300 & 80\\\hline
\end{tabular}
\caption{\label{Tab:ARPES_AF}Comparison of the temperature and photon energies of the different ARPES studies on undoped \sriro\ of which data is shown here.}
\end{center}
\end{table}

%%%%%%%%%%%%%%%%%%%%%%%%%%%%%%%%%%%

\section*{References}
\bibliography{Literature}

\begin{thebibliography}{10}

\bibitem{Gunnarsson1996}
O.~Gunnarsson, E.~Koch, and R.~M. Martin.
\newblock {Mott transition in degenerate Hubbard models: Application to doped
  fullerenes}.
\newblock {\em Phys. Rev. B}, 54:R11026--R11029, Oct 1996.

\bibitem{Rozenberg1997}
Marcelo~J. Rozenberg.
\newblock {Integer-filling metal-insulator transitions in the degenerate
  Hubbard model}.
\newblock {\em Phys. Rev. B}, 55:R4855--R4858, Feb 1997.

\bibitem{Imada1998}
M.~Imada, A.~Fujimori, and Y.~Tokura.
\newblock {Metal-insulator transitions}.
\newblock {\em Rev. Mod. Phys.}, 70:1039--1263, October 1998.

\bibitem{Pavarini2004}
E.~Pavarini, S.~Biermann, A.~Poteryaev, A.~I. Lichtenstein, A.~Georges, and
  O.~K. Andersen.
\newblock {Mott Transition and Suppression of Orbital Fluctuations in
  Orthorhombic $3{d}^{1}$ Perovskites}.
\newblock {\em Phys. Rev. Lett.}, 92:176403, Apr 2004.

\bibitem{Tomczak2014}
Jan~M. Tomczak, M.~Casula, T.~Miyake, and S.~Biermann.
\newblock {Asymmetry in band widening and quasiparticle lifetimes in
  ${\mathrm{SrVO}}_{3}$: Competition between screened exchange and local
  correlations from combined $GW$ and dynamical mean-field theory $GW +
  \mathrm{DMFT}$}.
\newblock {\em Phys. Rev. B}, 90:165138, Oct 2014.

\bibitem{Seth2017}
P.~{Seth}, O.~E. {Peil}, L.~{Pourovskii}, M.~{Betzinger}, C.~{Friedrich},
  O.~{Parcollet}, S.~{Biermann}, F.~{Aryasetiawan}, and A.~{Georges}.
\newblock {Renormalization of effective interactions in a negative charge
  transfer insulator}.
\newblock {\em Phys. Rev. B}, 96(20):205139, November 2017.

\bibitem{Lee2006}
P.~A. Lee, N.~Nagaosa, and X.-G. Wen.
\newblock {Doping a Mott insulator: Physics of high-temperature
  superconductivity}.
\newblock {\em Rev. Mod. Phys.}, 78:17--85, Jan 2006.

\bibitem{Scalapino2012}
D.~J. Scalapino.
\newblock {A common thread: The pairing interaction for unconventional
  superconductors}.
\newblock {\em Rev. Mod. Phys.}, 84:1383--1417, October 2012.

\bibitem{Alloul2014}
Henri A.
\newblock {What is the simplest model that captures the basic experimental
  facts of the physics of underdoped cuprates?}
\newblock {\em Comptes Rendus Physique}, 15(6):519 -- 524, 2014.

\bibitem{Andersen1995}
O.K. Andersen, A.I. Liechtenstein, O.~Jepsen, and F.~Paulsen.
\newblock {LDA energy bands, low-energy hamiltonians, $t^{\prime}$,
  $t^{\prime\prime}$, $t_{\perp}(\mathbf{k})$, and $J_{\perp}$}.
\newblock {\em Journal of Physics and Chemistry of Solids}, 56(12):1573 --
  1591, 1995.
\newblock Proceedings of the Conference on Spectroscopies in Novel
  Superconductors.

\bibitem{Hanke2010}
W.~Hanke, M.L. Kiesel, M.~Aichhorn, S.~Brehm, and E.~Arrigoni.
\newblock {The 3-band Hubbard-model versus the 1-band model for the high-Tc
  cuprates: Pairing dynamics, superconductivity and the ground-state phase
  diagram}.
\newblock {\em The European Physical Journal Special Topics}, 188(1):15--32,
  Oct 2010.

\bibitem{Kim2008}
B.~J. Kim, Hosub Jin, S.~J. Moon, J.-Y. Kim, B.-G. Park, C.~S. Leem, Jaejun Yu,
  T.~W. Noh, C.~Kim, S.-J. Oh, J.-H. Park, V.~Durairaj, G.~Cao, and
  E.~Rotenberg.
\newblock {Novel ${J}_{\mathrm{eff}}=1/2$ Mott State Induced by Relativistic
  Spin-Orbit Coupling in ${\mathrm{Sr}}_{2}{\mathrm{IrO}}_{4}$}.
\newblock {\em Phys. Rev. Lett.}, 101:076402, Aug 2008.

\bibitem{Martins2011}
C.~Martins, M.~Aichhorn, L.~Vaugier, and S.~Biermann.
\newblock {Reduced Effective Spin-Orbital Degeneracy and Spin-Orbital Ordering
  in Paramagnetic Transition-Metal Oxides:
  ${\mathrm{Sr}}_{2}{\mathrm{IrO}}_{4}$ versus
  ${\mathrm{Sr}}_{2}{\mathrm{RhO}}_{4}$}.
\newblock {\em Phys. Rev. Lett.}, 107:266404, Dec 2011.

\bibitem{Arita2012}
R.~Arita, J.~Kune\ifmmode~\check{s}\else \v{s}\fi{}, A.~V. Kozhevnikov, A.~G.
  Eguiluz, and M.~Imada.
\newblock {Ab initio Studies on the Interplay between Spin-Orbit Interaction
  and Coulomb Correlation in Sr${}_2$IrO${}_4$ and Ba${}_2$IrO${}_4$}.
\newblock {\em Phys. Rev. Lett.}, 108:086403, Feb 2012.

\bibitem{Martins2017}
C.~Martins, M.~Aichhorn, and S.~Biermann.
\newblock {Coulomb correlations in 4d and 5d oxides from first principles -- or
  how spin-orbit materials choose their effective orbital degeneracies}.
\newblock {\em Journal of Physics: Condensed Matter}, 29(26):263001, 2017.

\bibitem{Haskel2012}
D.~Haskel, G.~Fabbris, Mikhail Zhernenkov, P.~P. Kong, C.~Q. Jin, G.~Cao, and
  M.~van Veenendaal.
\newblock {Pressure Tuning of the Spin-Orbit Coupled Ground State in
  ${\mathrm{Sr}}_{2}{\mathrm{IrO}}_{4}$}.
\newblock {\em Phys. Rev. Lett.}, 109:027204, Jul 2012.

\bibitem{Fujiyama2012}
S.~Fujiyama, H.~Ohsumi, T.~Komesu, J.~Matsuno, B.~J. Kim, M.~Takata, T.~Arima,
  and H.~Takagi.
\newblock {Two-Dimensional Heisenberg Behavior of
  ${J}_{\mathrm{eff}}\mathbf{=}1/2$ Isospins in the Paramagnetic State of the
  Spin-Orbital Mott Insulator ${\mathrm{Sr}}_{2}{\mathrm{IrO}}_{4}$}.
\newblock {\em Phys. Rev. Lett.}, 108:247212, Jun 2012.

\bibitem{Kim2012}
J.~Kim, D.~Casa, M.~H. Upton, T.~Gog, Y.-J. Kim, J.~F. Mitchell, M.~van
  Veenendaal, M.~Daghofer, J.~van~den Brink, G.~Khaliullin, and B.~J. Kim.
\newblock {Magnetic Excitation Spectra of ${\mathrm{Sr}}_{2}{\mathrm{IrO}}_{4}$
  Probed by Resonant Inelastic X-Ray Scattering: Establishing Links to Cuprate
  Superconductors}.
\newblock {\em Phys. Rev. Lett.}, 108:177003, Apr 2012.

\bibitem{Fujiyama2014}
S.~Fujiyama, H.~Ohsumi, K.~Ohashi, D.~Hirai, B.~J. Kim, T.~Arima, M.~Takata,
  and H.~Takagi.
\newblock {Spin and Orbital Contributions to Magnetically Ordered Moments in
  $5d$ Layered Perovskite {S}r$_{2}${I}r{O}$_{4}$}.
\newblock {\em Phys. Rev. Lett.}, 112:016405, Jan. 2014.

\bibitem{Kim2014}
Y.~K. Kim, O.~Krupin, J.~D. Denlinger, A.~Bostwick, E.~Rotenberg, Q.~Zhao,
  J.~F. Mitchell, J.~W. Allen, and B.~J. Kim.
\newblock {Fermi arcs in a doped pseudospin-1/2 Heisenberg antiferromagnet}.
\newblock {\em Science}, 345(6193):187--190, 2014.

\bibitem{Wang2013}
Q.~Wang, Y.~Cao, J.~A. Waugh, S.~R. Park, T.~F. Qi, O.~B. Korneta, G.~Cao, and
  D.~S. Dessau.
\newblock {Dimensionality-controlled Mott transition and correlation effects in
  single-layer and bilayer perovskite iridates}.
\newblock {\em Phys. Rev. B}, 87:245109, Jun 2013.

\bibitem{Yamasaki2014}
A.~Yamasaki, S.~Tachibana, H.~Fujiwara, A.~Higashiya, A.~Irizawa, O.~Kirilmaz,
  F.~Pfaff, P.~Scheiderer, J.~Gabel, M.~Sing, T.~Muro, M.~Yabashi, K.~Tamasaku,
  H.~Sato, H.~Namatame, M.~Taniguchi, A.~Hloskovskyy, H.~Yoshida, H.~Okabe,
  M.~Isobe, J.~Akimitsu, W.~Drube, R.~Claessen, T.~Ishikawa, S.~Imada,
  A.~Sekiyama, and S.~Suga.
\newblock {Bulk nature of layered perovskite iridates beyond the Mott scenario:
  An approach from a bulk-sensitive photoemission study}.
\newblock {\em Phys. Rev. B}, 89:121111, Mar 2014.

\bibitem{Brouet2015}
V.~Brouet, J.~Mansart, L.~Perfetti, C.~Piovera, I.~Vobornik, P.~Le~F\`evre,
  F.~Bertran, S.~C. Riggs, M.~C. Shapiro, P.~Giraldo-Gallo, and I.~R. Fisher.
\newblock {Transfer of spectral weight across the gap of
  ${\mathrm{Sr}}_{2}{\mathrm{IrO}}_{4}$ induced by La doping}.
\newblock {\em Phys. Rev. B}, 92:081117, Aug 2015.

\bibitem{Torre2015}
A.~de~la Torre, S.~McKeown~Walker, F.~Y. Bruno, S.~Ricc\'o, Z.~Wang,
  I.~Gutierrez~Lezama, G.~Scheerer, G.~Giriat, D.~Jaccard, C.~Berthod, T.~K.
  Kim, M.~Hoesch, E.~C. Hunter, R.~S. Perry, A.~Tamai, and F.~Baumberger.
\newblock {Collapse of the Mott Gap and Emergence of a Nodal Liquid in Lightly
  Doped ${\mathrm{Sr}}_{2}{\mathrm{IrO}}_{4}$}.
\newblock {\em Phys. Rev. Lett.}, 115:176402, Oct 2015.

\bibitem{Liu2015}
Y.~Liu, L.~Yu, X.~Jia, J.~Zhao, H.~Weng, Y.~Peng, C.~Chen, Z.~Xie, D.~Mou,
  J.~He, X.~Liu, Y.~Feng, H.~Yi, L.~Zhao, G.~Liu, S.~He, X.~Dong, J.~Zhang,
  Z.~Xu, C.~Chen, G.~Cao, X.~Dai, Z.~Fang, and X.~J. Zhou.
\newblock {Anomalous High-Energy Waterfall-Like Electronic Structure in 5d
  Transition Metal Oxide Sr${}_2$IrO${}_4$ with a Strong Spin-Orbit Coupling}.
\newblock {\em Sci. Rep.}, 5:13036--, August 2015.

\bibitem{Nie2015}
Y.~F. Nie, P.~D.~C. King, C.~H. Kim, M.~Uchida, H.~I. Wei, B.~D. Faeth, J.~P.
  Ruf, J.~P.~C. Ruff, L.~Xie, X.~Pan, C.~J. Fennie, D.~G. Schlom, and K.~M.
  Shen.
\newblock {Interplay of Spin-Orbit Interactions, Dimensionality, and Octahedral
  Rotations in Semimetallic SrIrO${}_{3}$}.
\newblock {\em Phys. Rev. Lett.}, 114:016401, Jan 2015.

\bibitem{Piovera2016}
C.~Piovera, V.~Brouet, E.~Papalazarou, M.~Caputo, M.~Marsi, A.~Taleb-Ibrahimi,
  B.~J. Kim, and L.~Perfetti.
\newblock {Time-resolved photoemission of
  ${\mathrm{Sr}}_{2}{\mathrm{IrO}}_{4}$}.
\newblock {\em Phys. Rev. B}, 93:241114, Jun 2016.

\bibitem{Cao2016}
Y.~Cao, Q.~Wang, J.~A. Waugh, T.~J. Reber, H.~Li, X.~Zhou, S.~Parham, S.-R.
  Park, N.~C. Plumb, E.~Rotenberg, A.~Bostwick, J.~D. Denlinger, T.~Qi, M.~A.
  Hermele, G.~Cao, and D.~S. Dessau.
\newblock {Hallmarks of the Mott-metal crossover in the hole-doped
  pseudospin-1/2 Mott insulator Sr${}_2$IrO${}_4$}.
\newblock {\em Nature Communications}, 7:11367--, March 2016.

\bibitem{Kim2016}
Y.~K. Kim, N.~H. Sung, J.~D. Denlinger, and B.~J. Kim.
\newblock {Observation of a d-wave gap in electron-doped Sr${}_2$IrO${}_4$}.
\newblock {\em Nat Phys}, 12(1):37--41, January 2016.

\bibitem{Martins2018}
C.~Martins, B.~Lenz, L.~Perfetti, V.~Brouet, F.~Bertran, and S.~Biermann.
\newblock {Nonlocal Coulomb correlations in pure and electron-doped
  ${\mathrm{Sr}}_{2}{\mathrm{IrO}}_{4}$: Spectral functions, Fermi surface, and
  pseudo-gap-like spectral weight distributions from oriented cluster dynamical
  mean-field theory}.
\newblock {\em Phys. Rev. Materials}, 2:032001, Mar 2018.

\bibitem{Li2013}
Q.~Li, G.~Cao, S.~Okamoto, J.~Yi, W.~Lin, B.~C. Sales, J.~Yan, R.~Arita,
  J.~Kunes, A.~V. Kozhevnikov, A.~G. Eguiluz, M.~Imada, Z.~Gai, M.~Pan, and
  D.~G. Mandrus.
\newblock {Atomically resolved spectroscopic study of Sr${}_2$IrO${}_4$:
  Experiment and theory}.
\newblock {\em Scientific Reports}, 3:3073, Oct 2013.

\bibitem{Dai2014}
J.~Dai, E.~Calleja, G.~Cao, and K.~McElroy.
\newblock {Local density of states study of a spin-orbit-coupling induced Mott
  insulator ${\mathrm{Sr}}_{2}\mathrm{Ir}{\mathrm{O}}_{4}$}.
\newblock {\em Phys. Rev. B}, 90:041102, Jul 2014.

\bibitem{Moon2008}
S.~J. Moon, H.~Jin, K.~W. Kim, W.~S. Choi, Y.~S. Lee, J.~Yu, G.~Cao, A.~Sumi,
  H.~Funakubo, C.~Bernhard, and T.~W. Noh.
\newblock {Dimensionality-Controlled Insulator-Metal Transition and Correlated
  Metallic State in $5d$ Transition Metal Oxides
  ${\mathrm{Sr}}_{n+1}{\mathrm{Ir}}_{n}{\mathrm{O}}_{3n+1}$ ($n=1$, 2, and
  $\infty$)}.
\newblock {\em Phys. Rev. Lett.}, 101:226402, Nov 2008.

\bibitem{Moon2009}
S.~J. Moon, H.~Jin, W.~S. Choi, J.~S. Lee, S.~S.~A. Seo, J.~Yu, G.~Cao, T.~W.
  Noh, and Y.~S. Lee.
\newblock {Temperature dependence of the electronic structure of the
  $J_{eff}=\frac{1}{2}$ Mott insulator Sr${}_2$IrO${}_4$ studied by optical
  spectroscopy}.
\newblock {\em Phys. Rev. B}, 80:195110, Nov 2009.

\bibitem{Hsieh2012}
D.~Hsieh, F.~Mahmood, D.~H. Torchinsky, G.~Cao, and N.~Gedik.
\newblock {Observation of a metal-to-insulator transition with both
  Mott-Hubbard and Slater characteristics in Sr${}_{2}$IrO${}_{4}$ from
  time-resolved photocarrier dynamics}.
\newblock {\em Phys. Rev. B}, 86:035128, Jul 2012.

\bibitem{Zhao2016}
L.~Zhao, D.~H. Torchinsky, H.~Chu, V.~Ivanov, R.~Lifshitz, R.~Flint, T.~Qi,
  G.~Cao, and D.~Hsieh.
\newblock {Evidence of an odd-parity hidden order in a spin-orbit coupled
  correlated iridate}.
\newblock {\em Nat Phys}, 12(1):32--36, January 2016.

\bibitem{Proepper2016}
D.~Pr\"opper, A.~N. Yaresko, M.~H\"oppner, Y.~Matiks, Y.-L. Mathis,
  T.~Takayama, A.~Matsumoto, H.~Takagi, B.~Keimer, and A.~V. Boris.
\newblock {Optical anisotropy of the ${J}_{\mathrm{eff}}=1/2$ Mott insulator
  ${\mathrm{Sr}}_{2}{\mathrm{IrO}}_{4}$}.
\newblock {\em Phys. Rev. B}, 94:035158, Jul 2016.

\bibitem{Li2016}
Y.~Li, R.~D. Schaller, M.~Zhu, D.~A. Walko, J.~Kim, X.~Ke, L.~Miao, and Z.~Q.
  Mao.
\newblock {Strong lattice correlation of non-equilibrium quasiparticles in a
  pseudospin-1/2 Mott insulator Sr${}_2$IrO${}_4$}.
\newblock {\em Scientific Reports}, 6:19302, Jan 2016.

\bibitem{Chikara2009}
S.~Chikara, O.~Korneta, W.~P. Crummett, L.~E. DeLong, P.~Schlottmann, and
  G.~Cao.
\newblock {Giant magnetoelectric effect in the ${J}_{\mathrm{eff}}=\frac{1}{2}$
  Mott insulator Sr${}_2$IrO${}_4$}.
\newblock {\em Phys. Rev. B}, 80:140407, Oct 2009.

\bibitem{Ge2011}
M.~Ge, T.~F. Qi, O.~B. Korneta, D.~E. De~Long, P.~Schlottmann, W.~P. Crummett,
  and G.~Cao.
\newblock {Lattice-driven magnetoresistivity and metal-insulator transition in
  single-layered iridates}.
\newblock {\em Phys. Rev. B}, 84:100402, Sep 2011.

\bibitem{Pallecchi2016}
I.~Pallecchi, M.~T. Buscaglia, V.~Buscaglia, E.~Gilioli, G.~Lamura, F.~Telesio,
  M.R. Cimberle, and D.~Marr\'e.
\newblock {Thermoelectric behavior of Ruddlesden-Popper series iridates}.
\newblock {\em Journal of Physics: Condensed Matter}, 28(6):065601, 2016.

\bibitem{Klein2008}
Y.~Klein and I.~Terasaki.
\newblock {Insight on the electronic state of Sr${}_2$IrO${}_4$ revealed by
  cationic substitutions}.
\newblock {\em Journal of Physics: Condensed Matter}, 20(29):295201, 2008.

\bibitem{Kim2009}
B.~J. Kim, H.~Ohsumi, T.~Komesu, S.~Sakai, T.~Morita, H.~Takagi, and T.~Arima.
\newblock {Phase-Sensitive Observation of a Spin-Orbital Mott State in
  Sr${}_2$IrO${}_4$}.
\newblock {\em Science}, 323(5919):1329--1332, 2009.

\bibitem{Zhou2017}
Guotai Z., Xiaomin G., Xingming Y., Xiaoyang G., Kang W., Jin P., Fengming Z.,
  and X.~S. Wu.
\newblock {Structure, magnetism, and transport properties for Ca doping in
  Sr${}_2$IrO${}_4$}.
\newblock {\em AIP Advances}, 7(5):055823, 2017.

\bibitem{Zocco2014}
D~A Zocco, J~J Hamlin, B~D White, B~J Kim, J~R Jeffries, S~T Weir, Y~K Vohra,
  J~W Allen, and M~B Maple.
\newblock {Persistent non-metallic behavior in Sr${}_2$IrO${}_4$ and
  Sr${}_3$Ir${}_2$O${}_7$ at high pressures}.
\newblock {\em Journal of Physics: Condensed Matter}, 26(25):255603, 2014.

\bibitem{Fruchter2015}
L.~Fruchter, G.~Collin, D.~Colson, and V.~Brouet.
\newblock {Sr${}_2$IrO${}_4$ magnetic phase diagram from resistivity}.
\newblock {\em The European Physical Journal B}, 88(6):141, Jun 2015.

\bibitem{Zhang2013}
H.~Zhang, K.~Haule, and D.~Vanderbilt.
\newblock {Effective $J\mathbf{=}1/2$ Insulating State in Ruddlesden-Popper
  Iridates: An $\mathrm{LDA}\mathbf{+}\mathrm{DMFT}$ Study}.
\newblock {\em Phys. Rev. Lett.}, 111:246402, Dec 2013.

\bibitem{Hampel2015}
A.~Hampel, C.~Piefke, and F.~Lechermann.
\newblock {Low-energy model and electron-hole doping asymmetry of single-layer
  Ruddlesden-Popper iridates}.
\newblock {\em Phys. Rev. B}, 92:085141, Aug 2015.

\bibitem{Moutenet2018}
A.~Moutenet, A.~Georges, and M.~Ferrero.
\newblock {Pseudogap and electronic structure of electron-doped
  ${\mathrm{Sr}}_{2}{\mathrm{IrO}}_{4}$}.
\newblock {\em Phys. Rev. B}, 97:155109, Apr 2018.

\bibitem{Paerschke2017}
E.~M. P{\"a}rschke, K.~Wohlfeld, K.~Foyevtsova, and J.~van~den Brink.
\newblock {Correlation induced electron-hole asymmetry in quasi-
  two-dimensional iridates}.
\newblock {\em Nature Communications}, 8(1):686--, 2017.

\bibitem{Zhou2017b}
S.~Zhou, K.~Jiang, H.~Chen, and Z.~Wang.
\newblock {Correlation Effects and Hidden Spin-Orbit Entangled Electronic Order
  in Parent and Electron-Doped Iridates ${\mathrm{Sr}}_{2}{\mathrm{IrO}}_{4}$}.
\newblock {\em Phys. Rev. X}, 7:041018, Oct 2017.

\bibitem{Senthil2011}
F.~Wang and T.~Senthil.
\newblock {Twisted Hubbard Model for ${\mathrm{Sr}}_{2}{\mathrm{IrO}}_{4}$:
  Magnetism and Possible High Temperature Superconductivity}.
\newblock {\em Phys. Rev. Lett.}, 106:136402, Mar 2011.

\bibitem{Gretarsson2016}
H.~Gretarsson, N.~H. Sung, J.~Porras, J.~Bertinshaw, C.~Dietl, Jan A.~N. Bruin,
  A.~F. Bangura, Y.~K. Kim, R.~Dinnebier, Jungho Kim, A.~Al-Zein,
  M.~Moretti~Sala, M.~Krisch, M.~Le~Tacon, B.~Keimer, and B.~J. Kim.
\newblock {Persistent Paramagnons Deep in the Metallic Phase of
  Sr${}_{2-x}$La${}_{x}$IrO${}_{4}$}.
\newblock {\em Phys. Rev. Lett.}, 117:107001, Sep 2016.

\bibitem{Pincini2017}
D.~Pincini, J.~G. Vale, C.~Donnerer, A.~de~la Torre, E.~C. Hunter, R.~Perry,
  M.~Moretti~Sala, F.~Baumberger, and D.~F. McMorrow.
\newblock {Anisotropic exchange and spin-wave damping in pure and
  electron-doped ${\mathrm{Sr}}_{2}{\mathrm{IrO}}_{4}$}.
\newblock {\em Phys. Rev. B}, 96:075162, Aug 2017.

\bibitem{Headings2010}
N.~S. Headings, S.~M. Hayden, R.~Coldea, and T.~G. Perring.
\newblock {Anomalous High-Energy Spin Excitations in the High-${T}_{c}$
  Superconductor-Parent Antiferromagnet ${\mathrm{La}}_{2}{\mathrm{CuO}}_{4}$}.
\newblock {\em Phys. Rev. Lett.}, 105:247001, Dec 2010.

\bibitem{Ye2013}
F.~Ye, S.~Chi, B.~C. Chakoumakos, J.~A. Fernandez-Baca, T.~Qi, and G.~Cao.
\newblock {Magnetic and crystal structures of
  ${\mathrm{Sr}}_{2}{\mathrm{IrO}}_{4}$: A neutron diffraction study}.
\newblock {\em Phys. Rev. B}, 87:140406, Apr 2013.

\bibitem{Cosio-Castaneda2007}
C.~Cosio-Castaneda, G.~Tavizon, A.~Baeza, P.~de~la Mora, and R.~Escudero.
\newblock {Structure and magnetic properties of the weak ferromagnet
  Sr${}_{2-x}$La${}_x$IrO${}_4$}.
\newblock {\em Journal of Physics: Condensed Matter}, 19(44):446210, 2007.

\bibitem{Gatimu2012}
A.~J. Gatimu, R.~Berthelot, S.~Muir, A.~W. Sleight, and M.~A. Subramanian.
\newblock {Synthesis and characterization of Sr${}_2$Ir${}_{1-x}$M${}_x$O${}_4$
  ($M=Ti, Fe, Co$) solid solutions}.
\newblock {\em Journal of Solid State Chemistry}, 190:257 -- 263, 2012.

\bibitem{Korneta2010}
O.~B. Korneta, T.~Qi, S.~Chikara, S.~Parkin, L.~E. De~Long, P.~Schlottmann, and
  G.~Cao.
\newblock {Electron-doped Sr${}_2$IrO${}_{4-\delta}$ ($0\le\delta\le 0.04$):
  Evolution of a disordered J${}_{\mathrm{eff}}=\frac{1}{2}$ Mott insulator
  into an exotic metallic state}.
\newblock {\em Phys. Rev. B}, 82:115117, Sep 2010.

\bibitem{Alloul1989}
H.~Alloul, T.~Ohno, and P.~Mendels.
\newblock {${}^{89}\mathrm{Y}$ NMR Evidence for a Fermi-Liquid Behavior in
  YBa${}_2$Cu${}_3$O${}_{6+x}$}.
\newblock {\em Phys. Rev. Lett.}, 63:1700--1703, Oct 1989.

\bibitem{Taillefer2010}
L.~Taillefer.
\newblock {Scattering and Pairing in Cuprate Superconductors}.
\newblock {\em Annual Review of Condensed Matter Physics}, 1(1):51--70, 2010.

\bibitem{Sacuto2013}
A.~Sacuto, Y.~Gallais, M.~Cazayous, M.-A. M\'easson, G.~D. Gu, and D.~Colson.
\newblock {New insights into the phase diagram of the copper oxide
  superconductors from electronic Raman scattering}.
\newblock {\em Reports on Progress in Physics}, 76(2):022502, 2013.

\bibitem{Kordyuk2015}
A.~A. Kordyuk.
\newblock {Pseudogap from ARPES experiment: Three gaps in cuprates and
  topological superconductivity (Review Article)}.
\newblock {\em Low Temperature Physics}, 41(5):319--341, 2015.

\bibitem{Proust2018}
C.~{Proust} and L.~{Taillefer}.
\newblock {The remarkable underlying ground states of cuprate superconductors}.
\newblock {\em ArXiv e-prints}, July 2018.

\bibitem{Phillips2003}
J.~C. Phillips, A.~Saxena, and A.~R. Bishop.
\newblock {Pseudogaps, dopants, and strong disorder in cuprate high-temperature
  superconductors}.
\newblock {\em Reports on Progress in Physics}, 66(12):2111, 2003.

\bibitem{vanRoekeghem2016}
A.~van Roekeghem, P.~Richard, H.~Ding, and S.~Biermann.
\newblock {Spectral properties of transition metal pnictides and chalcogenides:
  Angle-resolved photoemission spectroscopy and dynamical mean-field theory}.
\newblock {\em Comptes Rendus Physique}, 17:140--163, 2016.

\bibitem{Georges1996}
A.~Georges, G.~Kotliar, W.~Krauth, and M.~J. Rozenberg.
\newblock {Dynamical mean-field theory of strongly correlated fermion systems
  and the limit of infinite dimensions}.
\newblock {\em Rev. Mod. Phys.}, 68:13--125, January 1996.

\bibitem{Kotliar2004}
G.~Kotliar and D.~Vollhardt.
\newblock {Strongly Correlated Materials: Insights From Dynamical Mean-Field
  Theory}.
\newblock {\em Physics Today}, 57(3):53--59, 2004.

\bibitem{Anisimov1997}
V.~I. Anisimov, A.~I. Poteryaev, M.~A. Korotin, A.~O. Anokhin, and G.~Kotliar.
\newblock {First-principles calculations of the electronic structure and
  spectra of strongly correlated systems: dynamical mean-field theory}.
\newblock {\em Journal of Physics: Condensed Matter}, 9(35):7359, 1997.

\bibitem{Lichtenstein1998}
A.~I. Lichtenstein and M.~I. Katsnelson.
\newblock {Ab initio calculations of quasiparticle band structure in correlated
  systems: LDA++ approach}.
\newblock {\em Phys. Rev. B}, 57:6884--6895, Mar 1998.

\bibitem{Biermann2014}
S.~Biermann.
\newblock {\em {Dynamical Mean Field Theory-Based Electronic Structure
  Calculations for Correlated Materials}}, pages 303--345.
\newblock Springer Berlin Heidelberg, 2014.

\bibitem{Biermann2005}
S.~Biermann, A.~Poteryaev, A.~I. Lichtenstein, and A.~Georges.
\newblock {Dynamical Singlets and Correlation-Assisted Peierls Transition in
  ${\mathrm{V}\mathrm{O}}_{2}$}.
\newblock {\em Phys. Rev. Lett.}, 94:026404, Jan 2005.

\bibitem{Kotliar2006}
G.~Kotliar, S.~Y. Savrasov, K.~Haule, V.~S. Oudovenko, O.~Parcollet, and C.~A.
  Marianetti.
\newblock {Electronic structure calculations with dynamical mean-field theory}.
\newblock {\em Rev. Mod. Phys.}, 78:865--951, August 2006.

\bibitem{Ma2014}
J.-Z. Ma, A.~van Roekeghem, P.~Richard, Z.-H. Liu, H.~Miao, L.-K. Zeng, N.~Xu,
  M.~Shi, C.~Cao, J.-B. He, G.-F. Chen, Y.-L. Sun, G.-H. Cao, S.-C. Wang,
  S.~Biermann, T.~Qian, and H.~Ding.
\newblock {Correlation-Induced Self-Doping in the Iron-Pnictide Superconductor
  ${\mathrm{Ba}}_{2}{\mathrm{Ti}}_{2}{\mathrm{Fe}}_{2}{\mathrm{As}}_{4}\mathrm{O}$}.
\newblock {\em Phys. Rev. Lett.}, 113:266407, Dec 2014.

\bibitem{Nowadnick2015}
E.~A. Nowadnick, J.~P. Ruf, H.~Park, P.~D.~C. King, D.~G. Schlom, K.~M. Shen,
  and A.~J. Millis.
\newblock {Quantifying electronic correlation strength in a complex oxide: A
  combined DMFT and ARPES study of LaNiO${}_{3}$}.
\newblock {\em Phys. Rev. B}, 92:245109, Dec 2015.

\bibitem{Hausoel2017}
A.~Hausoel, M.~Karolak, E.~Sasioglu, A.~Lichtenstein, K.~Held, A.~Katanin,
  A.~Toschi, and G.~Sangiovanni.
\newblock {Local magnetic moments in iron and nickel at ambient and Earth's
  core conditions}.
\newblock {\em Nature Communications}, 8:16062--, July 2017.

\bibitem{Hariki2017}
Atsushi Hariki, Andreas Hausoel, Giorgio Sangiovanni, and Jan
  Kune\ifmmode~\check{s}\else \v{s}\fi{}.
\newblock {DFT+DMFT study on soft moment magnetism and covalent bonding in
  SrRu${}_{2}$O${}_{6}$}.
\newblock {\em Phys. Rev. B}, 96:155135, Oct 2017.

\bibitem{Mandal2018}
Subhasish Mandal, R.~E. Cohen, and K.~Haule.
\newblock {Valence and spin fluctuations in the Mn-doped ferroelectric
  ${\mathrm{BaTiO}}_{3}$}.
\newblock {\em Phys. Rev. B}, 98:075155, Aug 2018.

\bibitem{Martins2010}
Cyril Martins.
\newblock {\em {Couplage Spin-Orbite et Interaction de Coulomb dans l'Iridate
  de Strontium Sr${}_2$IrO${}_4$}}.
\newblock PhD thesis, Ecole Polytechnique, Palaiseau, 2010.

\bibitem{Maier2005}
T.~Maier, M.~Jarrell, T.~Pruschke, and M.~H. Hettler.
\newblock Quantum cluster theories.
\newblock {\em Rev. Mod. Phys.}, 77:1027--1080, October 2005.

\bibitem{Aryasetiawan2004}
F.~Aryasetiawan, M.~Imada, A.~Georges, G.~Kotliar, S.~Biermann, and A.~I.
  Lichtenstein.
\newblock {Frequency-dependent local interactions and low-energy effective
  models from electronic structure calculations}.
\newblock {\em Phys. Rev. B}, 70:195104, Nov 2004.

\bibitem{Backes2019}
S.~Backes, H.~Jiang, and et~al.
\newblock unpublished.
\newblock 2019.

\bibitem{Hubbard1963}
J.~Hubbard.
\newblock {Electron correlations in narrow energy bands}.
\newblock {\em Proceedings of the Royal Society of London. Series A.
  Mathematical and Physical Sciences}, 276(1365):238--257, 1963.

\bibitem{Locht2017}
Inka L.~M. Locht.
\newblock {\em {Theoretical methods for the electronic structure and magnetism
  of strongly correlated materials}}.
\newblock PhD thesis, Uppsala University, Materials Theory, 2017.

\bibitem{Tomczak2007}
J.~M. Tomczak.
\newblock {\em {Propri\'et\'es spectrales et optiques des mat\'eriaux fortement
  corr\'el\'es}}.
\newblock PhD thesis, Ecole Polytechnique, Palaiseau, 2007.

\bibitem{Moser2014}
S.~Moser, L.~Moreschini, A.~Ebrahimi, B.~Dalla Piazza, M.~Isobe, H.~Okabe,
  J.~Akimitsu, V.~V. Mazurenko, K.~S. Kim, A.~Bostwick, E.~Rotenberg, J.~Chang,
  H.~M. R{\o}nnow, and M.~Grioni.
\newblock {The electronic structure of the high-symmetry perovskite iridate
  Ba${}_2$IrO${}_4$}.
\newblock {\em New Journal of Physics}, 16(1):013008, 2014.

\bibitem{Louat2018b}
Alex Louat.
\newblock {\em {\'Etudes spectroscopiques des nouveaux \'etats \'electroniques
  induits par fort couplage spin-orbite dans les iridates}}.
\newblock PhD thesis, Universit\'e Paris-Saclay, Orsay, 2018.

\bibitem{Louat2018}
A.~Louat, F.~Bert, F.~Bertran, P.~Le~F\`evre, J.~Rault, S.~Biermann,
  C.~Martins, B.~Lenz, and V.~Brouet.
\newblock {Orbital character and coherence energy scale defined by ARPES in
  Rh-doped Sr$_2$IrO$_4$}.
\newblock {\em in preparation}, 2018.

\bibitem{Dhital2013}
C.~Dhital, T.~Hogan, Z.~Yamani, C.~de~la Cruz, X.~Chen, S.~Khadka, Z.~Ren, and
  S.~D. Wilson.
\newblock {Neutron scattering study of correlated phase behavior in
  Sr${}_{2}$IrO${}_{4}$}.
\newblock {\em Phys. Rev. B}, 87:144405, Apr 2013.

\bibitem{Potthoff2003}
M.~Potthoff.
\newblock {Self-energy-functional approach to systems of correlated electrons}.
\newblock {\em The European Physical Journal B - Condensed Matter and Complex
  Systems}, 32:429--436, 2003.

\bibitem{Dahnken2004}
C.~Dahnken, M.~Aichhorn, W.~Hanke, E.~Arrigoni, and M.~Potthoff.
\newblock {Variational cluster approach to spontaneous symmetry breaking: The
  itinerant antiferromagnet in two dimensions}.
\newblock {\em Phys. Rev. B}, 70:245110, 2004.

\bibitem{Terashima2017}
K.~Terashima, M.~Sunagawa, H.~Fujiwara, T.~Fukura, M.~Fujii, K.~Okada,
  K.~Horigane, K.~Kobayashi, R.~Horie, J.~Akimitsu, E.~Golias, D.~Marchenko,
  A.~Varykhalov, N.~L. Saini, T.~Wakita, Y.~Muraoka, and T.~Yokoya.
\newblock {Evolution of the remnant Fermi-surface state in the lightly doped
  correlated spin-orbit insulator Sr${}_{2-x}$La${}_x$IrO${}_{4}$}.
\newblock {\em Phys. Rev. B}, 96:041106, Jul 2017.

\bibitem{Wallauer2015}
R.~Wallauer, S.~Sanna, E.~Lahoud, P.~Carretta, and A.~Kanigel.
\newblock {Sensitivity of angle-resolved photoemission to short-range
  antiferromagnetic correlations}.
\newblock {\em Phys. Rev. B}, 91:245149, Jun 2015.

\bibitem{Ayral2013}
Thomas Ayral, Silke Biermann, and Philipp Werner.
\newblock {Screening and nonlocal correlations in the extended Hubbard model
  from self-consistent combined $GW$ and dynamical mean field theory}.
\newblock {\em Phys. Rev. B}, 87:125149, Mar 2013.

\bibitem{Jeong2017}
J.~Jeong, Y.~Sidis, A.~Louat, V.~Brouet, and P.~Bourges.
\newblock {Time-reversal symmetry breaking hidden order in
  Sr${}_2$(Ir,Rh)O${}_4$}.
\newblock {\em Nature Communications}, 8:15119, April 2017.

\bibitem{Jin2009}
H.~Jin, H.~Jeong, T.~Ozaki, and J.~Yu.
\newblock {Anisotropic exchange interactions of spin-orbit-integrated states in
  Sr${}_{2}$IrO${}_{4}$}.
\newblock {\em Phys. Rev. B}, 80:075112, Aug 2009.

\bibitem{Carter2013}
J.-M. Carter, V.~Shankar, and H.-Y. Kee.
\newblock {Theory of metal-insulator transition in the family of perovskite
  iridium oxides}.
\newblock {\em Phys. Rev. B}, 88:035111, Jul 2013.

\end{thebibliography}
\end{document}